\documentclass[prd,superscriptaddress,twocolumn]{revtex4}

\usepackage{mathrsfs,amsmath}
\usepackage{eurosym}
\usepackage{amssymb}
\usepackage{bbold}

\usepackage{latexsym,epsfig,epstopdf}
\usepackage{amssymb,amsmath,amsthm}

\usepackage{times}
\usepackage{color}

\usepackage{lmodern}
\usepackage[utf8]{inputenc}

\newcommand{\ket}[1]{\left | #1 \right\rangle}
\newcommand{\bra}[1]{\left \langle #1 \right |}

\newcommand{\braket}[2]{\left\langle #1|#2\right\rangle}
\newcommand{\proj}[1]{\ket{#1}\bra{#1}}

\usepackage{graphicx}
\usepackage{color}
\usepackage{braket}
\usepackage[mathscr]{euscript} 
\usepackage{perpage} 
\usepackage{slashed}

\usepackage{enumitem}

\def\be{\begin{equation}}
\def\ee{\end{equation}}
\def\bea{\begin{eqnarray}}
\def\eea{\end{eqnarray}}

\usepackage{calc}
\newsavebox\CBox
\newcommand\hcancel[2][0.5pt]{%
  \ifmmode\sbox\CBox{$#2$}\else\sbox\CBox{#2}\fi%
  \makebox[0pt][l]{\usebox\CBox}%
  \rule[0.5\ht\CBox-#1/2]{\wd\CBox}{#1}}

\begin{document}

\title{Fractal properties of particle paths due to generalised uncertainty relations} 

\author{Matthew J. Lake}
\email{matthewjlake@narit.or.th}
\affiliation{National Astronomical Research Institute of Thailand, \\ 260 Moo 4, T. Donkaew,  A. Maerim, Chiang Mai 50180, Thailand}
\affiliation{Department of Physics, Faculty of Science, Chiang Mai University, \\ 239 Huaykaew Road, T. Suthep, A. Muang, Chiang Mai 50200, Thailand}
\affiliation{School of Physics, Sun Yat-Sen University, Guangzhou 510275, China}
\affiliation{Department of Physics, Babe\c s-Bolyai University, \\ Mihail Kog\u alniceanu Street 1, 400084 Cluj-Napoca, Transylvania, Romania}

\begin{abstract}
We determine the Hausdorff dimension of a particle path, $D_{\rm H}$, in the recently proposed `smeared space' model of quantum geometry. 
The model introduces additional degrees of freedom to describe the quantum state of the background and gives rise to both the generalised uncertainty principle (GUP) and extended uncertainty principle (EUP) without introducing modified commutation relations. 
We compare our results to previous studies of the Hausdorff dimension in GUP models based on modified commutators and show that the minimum length enters the relevant formulae in a different way. 
We then determine the Hausdorff dimension of the particle path in smeared momentum space, $\tilde{D}_{\rm H}$, and show that the minimum momentum is dual to the minimum length. 
For sufficiently coarse grained paths, $D_{\rm H} = \tilde{D}_{\rm H} = 2$, as in canonical quantum mechanics. 
However, as the resolutions approach the minimum scales, the dimensions of the paths in each representation differ, in contrast to their counterparts in the canonical theory. 
The GUP-induced corrections increase $D_{\rm H}$ whereas the EUP-induced corrections decrease $\tilde{D}_{\rm H}$, relative to their canonical values, and the extremal case corresponds to $D_{\rm H} = 3$, $\tilde{D}_{\rm H} = 1$. 
These results show that the GUP and the EUP affect the fractal properties of the particle path in fundamentally different, yet complimentary, ways. 
\end{abstract}

\maketitle


\tableofcontents

\section{Introduction} \label{Sec.1}

In canonical quantum mechanics (QM) the observed path of a particle is everywhere continuous but nowhere differentiable, as first noted by Feynman and Hibbs \cite{Feynman_Hibbs}. 
Due to Heisenberg's uncertainty principle (HUP) the distance travelled by the particle in a fixed time $\Delta t$ depends on the resolution of the detecting apparatus, $\Delta x$, so that the total path length $l(\Delta x)$ is resolution-dependent and diverges as $\Delta x \rightarrow 0$. 
These properties are shared by fractal curves \cite{Falconer} and the systematic study of the fractal properties of particle paths in canonical QM was inaugurated by Abbott and Wise \cite{Abbott:1979bh}. 
They showed that a modified definition of `length' known as the Hausdorff length, originally developed to analyse classical fractals  \cite{Falconer}, can be applied to QM paths. 

The Hausdorff length, $L_{\rm H}$, is defined as
\begin{eqnarray} \label{L_H}
L_{\rm H} = l \, . \, (\Delta x)^{D_{\rm H}-1} \, , 
\end{eqnarray}
where $l = l(\Delta x)$ is the resolution-dependent path length and the Hausdorff dimension $D_{\rm H}$ is chosen so that $L_{\rm H}$ is independent of $\Delta x$. 
For classical particle trajectories, $D_{\rm H} = 1$, but for $1 < D_{\rm H} \leq d$, where $d$ is the topological dimension of the background space, the Hausdorff `length' is no longer a true length. 
When $D_{\rm H} = 2,3, \dots d$, $L_{\rm H}$ has the dimensions of an area, volume, or $d$-dimensional hypervolume, respectively. 
This corresponds to the scenario in which the small-scale kinks in the particle path become so dense that the path effectively fills an $n$-dimensional hypersurface ($n \leq d$) within the $d$-dimensional space. 

In their pioneering work, Abbot and Wise considered resolving the path of a free particle in canonical QM, in three spatial dimensions, by performing a series of position measurements of accuracy $\Delta x$, separated by time-intervals $\Delta t$. 
They showed that, for $\Delta t \gg 4m (\Delta x)^2/\hbar$, the particle path becomes self-similar, i.e., fractal, with $D_{\rm H} = 2$ \cite{Abbott:1979bh}. 
However, it is important to note that this result holds in any number of topological dimensions. 
In other words, just as the classical trajectory of a particle is one-dimensional, regardless of the number of dimensions it propagates in, its quantum mechanical path always has Hausdorff dimension 2, irrespective of the dimensionality of the background geometry.   

In recent years, many studies of phenomenological quantum gravity have been conducted, in various spacetime dimensions. 
Most phenomenological models incorporate a minimum length scale, assumed to be of the order of the Planck length, while some also include a minimum momentum. (See \cite{Hossenfelder:2012jw} for a review.) 
Based on gedanken experiment arguments, it is believed that introducing minimum length and momentum scales alters the canonical HUP, giving rise to generalised uncertainty relations (GURs) \cite{Garay:1994en,Adler:1999bu,Scardigli:1999jh,Bolen:2004sq,Park:2007az,Bambi:2007ty}. 
In three spatial dimensions, the presence of a minimum length, $l_{\rm Pl} = \sqrt{\hbar G/c^3}$, gives rise to the GUP
\begin{eqnarray} \label{GUP}
\Delta x^i \gtrsim \frac{\hbar}{2 \Delta p_j} \delta^{i}{}_{j}\left[1 + \alpha_0 \frac{2G}{c^3}(\Delta p_j)^2\right] \, ,
\end{eqnarray}
where $\alpha_0$ is a numerical constant of order unity \cite{Adler:1999bu,Scardigli:1999jh}. 
For $d > 3$, $l_{\rm Pl}$ must be replaced with the $D$-dimensional Planck length, $l^{(D)}_{\rm Pl} = (\hbar G_{\rm D}/c^3)^{\frac{1}{D-2}}$, where $G_{\rm D}$ is the $D$-dimensional Newton's constant and $D = d+1$ is the number of spacetime dimensions \cite{Lake_Carr_2018}, but the algebraic form of the GUP remains independent of the topological dimension of the background. 

The presence of a minimum momentum, $m_{\rm dS}c = \hbar\sqrt{\Lambda/3}$, gives rise to the EUP 
\begin{eqnarray} \label{EUP}
\Delta p_j \gtrsim \frac{\hbar}{2 \Delta x^i} \delta^{i}{}_{j}\left[1 + 2\eta_0 \Lambda (\Delta x^i)^2\right] \, , 
\end{eqnarray}
where $\eta_0$ is of order unity \cite{Bolen:2004sq,Park:2007az,Bambi:2007ty}. 
Here, the parameter $\Lambda$ may be identified with the cosmological constant so that the de Sitter momentum, $m_{\rm dS}c$, is the momentum of a particle whose de Broglie wavelength is of the order of the present day radius of the universe \cite{GR_book}. 
In higher dimensions, the $D$-dimensional de Sitter momentum is $m_{\rm dS}^{(D)}c = \hbar\sqrt{\Lambda_{\rm D}/(D-1)}$, but the algebraic form of the EUP remains unaffected by the  dimensionality of the spacetime. 

Introducing both minimum length and minimum momentum scales yields the extended generalised uncertainty principle (EGUP), 
\begin{eqnarray} \label{EGUP}
\Delta x^i \Delta p_j \gtrsim \frac{\hbar}{2} \delta^{i}{}_{j} [1 + \alpha (\Delta p_j)^2 + \eta (\Delta x^i)^2] \, , 
\end{eqnarray}
where $\alpha$ and $\eta$ are dimensionful constants \cite{Bolen:2004sq,Park:2007az,Bambi:2007ty}. 
These are obtained by taking either $\alpha (\Delta p_j)^2$ or $\eta (\Delta x^i)^2$ as the subdominant term in Eq. (\ref{EGUP}) and comparing the different limits with Eqs. (\ref{GUP}) and (\ref{EUP}), respectively. 

Given the current research interest in quantum gravity it is natural to try to extend the work of Abbott and Wise to include Planck-scale effects using various phenomenological models. 
Naively, we may expect Planck-scale fluctuations of the background geometry to induce additional fluctuations in the particle path, increasing the Hausdorff dimension to $D_{\rm H} > 2$. 
The presence of a minimum length should also introduce an absolute lower bound for the resolution scale $\Delta x$ and, hence, an absolute upper bound for the total path length traversed in any time period $\Delta t$. 
In addition, the existence of a minimum length implies the existence of a minimum volume, $\sim (l_{\rm Pl}^{(D)})^{D-1}$, so that the GUP-modified Hausdorff dimension may, in principle, depend on the topological dimension of the space, $d = D-1$, in contrast to its canonical QM counterpart. 
As we will show in this paper, whether this is the case, or not, depends on which GUP model we choose. 

Until recently, all GUR models were assumed to arise from modified commutation relations \cite{Tawfik:2014zca,Tawfik:2015rva}, which, in turn, were derived from modified phase space volumes. 
In \cite{Kempf:1996ss}, Kempf, Mangano and Mann (KMM) showed that the GUP (\ref{GUP}) can be obtained rigorously, from the quantum formalism, by introducing the modified momentum space volume $(1 + \alpha \bold{p}^2){\rm d}^{d}p$. 
However, in this case, the position space representation of the theory is not well defined. 
Similarly, the EUP (\ref{EUP}) is obtained by introducing the modified position space volume $(1 + \eta \bold{x}^2){\rm d}^{d}x$, but, in this case, the momentum space representation is not well defined. 

In order to obtain the EGUP (\ref{EGUP}) from an appropriate modified commutator both the canonical position and momentum space representations must be abandoned and a generalised Bargman-Fock representation introduced \cite{Kempf_Bargmann_Fock}. 
Theoretically, this is not problematic, but it may also be shown that GURs based on modified commutators suffer from a number of unresolved pathologies, including violation of the equivalence principle, the reference frame-dependence of the `minimum' length, and the so called soccer ball problem for multi-particle states, among others \cite{Hossenfelder:2012jw,Tawfik:2014zca,Tawfik:2015rva}. 
The latter arises from the necessity of constructing a nonlinear addition law for the modified momenta but an ingenious resolution of this problem was proposed by Amelino-Camelia in \cite{Amelino-Camelia:2014gga}. 
Unfortunately, this is not applicable to many minimum-length models
\footnote{In Amelino-Camelia's approach, the generalised momentum operators, of a given modified commutator model, are considered as the generators of `generalised' spatial translations, by definition. 
This requires the unitary operator $\hat{\mathcal{U}}({\bf X}') := \exp[(i/\hbar){\bf X}'.\hat{{\bf P}}]$ to leave the modified $[\hat{X}^{i},\hat{P}_{j}]$, $[\hat{X}^{i},\hat{X}^{j}]$ and $[\hat{P}_{i},\hat{P}_{j}]$ algebras, as well as the multi-particle Hamiltonian of the model, invariant. 
If these invariances hold, in a given model, then the corresponding Noether charge for an $N$-particle state is represented by the operator $\hat{{\bf P}}_{\rm Total} := \sum_{i=1}^{N}\hat{{\bf P}_{i}}$, where $[\hat{{\bf P}}_{\rm Total} ,\hat{\mathcal{H}}] = 0$. 
In this scenario, the usual law of linear momentum addition still holds for multi-particle states but a different nonlinear addition law, which is derived ultimately from the notion of spatial locality, holds for transfers of momentum between individual particles, due to the interactions specified by $\hat{\mathcal{H}}$ \cite{Amelino-Camelia:2014gga}. 
\\ 
Unfortunately for GUP models, in the example system considered in \cite{Amelino-Camelia:2014gga}, the definition of the generalised spatial translations that is required to maintain the linear addition law also requires the relation $[\hat{X}_2,\hat{P}_2] = 0$ to hold. 
In this case, there is no Heisenberg uncertainty principle, let alone a GUP, even though a minimum length-scale $l$ still appears in the model via the position-position commutator $[\hat{X}_1,\hat{X}_2] = il\hat{X}_1$ .
\\ 
This illustrates a more general point: it is by no means certain that a particular modified momentum operator, corresponding to a particular modification of the canonical Heisenberg algebra, and, hence, a particular form of the GUP, is compatible with a linear addition law derived via Amelino-Camelia's procedure. 
In fact, we may consider applying this procedure to any prospective GUP model based on modified commutators and using it to rule out ones which give rise to inconsistencies. 
\\ 
In summary, although this procedure represents a useful criterion for a physically viable theory, it is clear that arbitrary deformations of the canonical algebras are not consistent with the existence of a linear momentum addition law and that further work is required to figure out which ones truly suffer from a soccer ball problem and which ones do not. 
Though some GUP models {\it may} be free from this pathology, it may be considered as likely that a great many are still afflicted by it.}.

Furthermore, although Abbott and Wise did not consider the fractal properties of the particle path in momentum space, directly, it is straightforward to show that in canonical QM the momentum space Hausdorff length can be defined as $\tilde{L}_{\rm H} = \tilde{l} \, . \, (\Delta p)^{\tilde{D}_{\rm H}-1}$, where $\tilde{l} = \tilde{l}(\Delta p)$, and that $\tilde{D}_{\rm H} = D_{\rm H} = 2$ when $\Delta t$ is sufficiently large. (See Sec. \ref{Sec.2.1}.) 
However, using traditional GUR models, we may investigate either the effects of a minimum length, using the GUP, or the effects of a minimum momentum, using the EUP, but not both using the EGUP
\footnote{In fact, it may be possible to analyse the effects of both minimum length and minimum momentum using the Bargmann-Fock representation of the EGUP, but this is by no means clear. Due to various technical problems with this approach, it is not currently known whether the theory is, in general, free from UV divergences \cite{Kempf_Bargmann_Fock}.}. 
In this work we use an alternative model of the EGUP, which successfully evades the pathologies of existing models based on modified commutators \cite{Lake:2018zeg,Lake:2019oaz,Lake:2019nmn,Lake:2021beh,Lake2020-1,Lake2020-2}, and which also allows us to the analyse the effects, on the fractal properties of a particle path, of both minimum length and momentum scales. 

The structure of this paper is as follows. 
In Sec. \ref{Sec.2} we review, and extend, previous work on the Hausdorff dimension of particle paths in canonical QM \cite{Abbott:1979bh} and in minimum-length models based on modified phase space volumes \cite{Nicolini:2010dj,NiednerThesis}. 
Sec. \ref{Sec.2.1} generalises the results of Abbott and Wise \cite{Abbott:1979bh} to arbitrary dimensions and includes new results in both the position and momentum space representations. 
Sec. \ref{Sec.2.2} reviews the work of Nicolini and Niedner \cite{Nicolini:2010dj,NiednerThesis} who considered a GUP model based on a modified momentum space volume, similar to that defined in the KMM model \cite{Kempf:1996ss}, and discusses a potential loophole in their analysis. 
Crucially, in their work, $D_{\rm H}$ was found to depend on $d$, the topological dimension of the background space. 
In Sec. \ref{Sec.3}, we review the smeared space model. 
Sec. \ref{Sec.3.1} reviews the basic formalism and a number of important differences between smeared space and previous minimum-length models, that are especially relevant for the analysis of the Hausdorff dimension, are highlighted in Sec. \ref{Sec.3.2}. 
In Sec. \ref{Sec.4}, we derive our main results. 
The fractal properties of paths obeying the smeared space GUP are investigated in Sec. \ref{Sec.4.1} and the Hausdorff dimension of the paths in the position space representation, $D_{\rm H}$, is determined. 
The fractal properties of particle paths obeying the smeared space EUP are investigated in Sec. \ref{Sec.4.1} and the Hausdorff dimension of the path in the momentum space representation, $\tilde{D}_{\rm H}$, is derived. 
Finally, in Sec. \ref{Sec.4.3}, we consider the implications of our model for the fractal properties of the dark energy field. 
Sec. \ref{Sec.5} contains a summary of our main conclusions and a brief discussion of prospects for future work. 

\section{The fractal properties of particle paths in canonical quantum mechanics and previous minimum-length models} \label{Sec.2}

In this section, we review and extend previous work on the Hausdorff dimension of particle paths in canonical QM \cite{Abbott:1979bh} and in models with a minimum length \cite{Nicolini:2010dj,NiednerThesis}. 

\subsection{Canonical quantum mechanics} \label{Sec.2.1} 

In \cite{Abbott:1979bh}, Abbott and Wise considered canonical quantum particles propagating in three spatial dimensions. 
Here, we give an outline of their work, but modify their original calculations slightly, in order to generalise them to an arbitrary number of dimensions, $d$. 
Our purpose is to show, explicitly, that the dimensionality of the background space has no effect on their main result, i.e., that the Hausdorff dimension of the path of a free particle is $D_{\rm H} = 2$, for any $d$. 

To this end, we consider resolving the path by performing a series of position measurements, each with resolution $(\Delta x)^d$, separated by fixed time intervals of $\Delta t$. 
The average distance traversed in a single time interval is 
\begin{eqnarray} \label{<Delta_l>-1}
\braket{\Delta l}_{\psi} = \braket{\psi | \hat{U}^{\dagger}(\Delta t) |\hat{\bold{x}}| \hat{U}(\Delta t) | \psi} = \int |\psi_{\Delta t}(\bold{x})|^2 |\bold{x}| {\rm d}^dx \, ,
\end{eqnarray}
where 
\begin{eqnarray} \label{U(t)}
\hat{U}(t) = \exp\left[-\frac{\hat{\bold{p}}^2}{2m} t\right]
\end{eqnarray}
is the time-evolution operator and $\psi_{\Delta t}(\bold{x}) = \braket{\bold{x} | \hat{U}(\Delta t) | \psi}$ is the state at $\Delta t$. 
For simplicity, we assume that each measurement collapses the wave function to a $d$-dimensional Gaussian, of width $\Delta x$ in each Cartesian direction, and that the initial wave function $\psi_0(\bold{x})$ is prepared in a similar way, so that
\begin{eqnarray} \label{psi_0}
|\psi_{0}(\bold{x})|^2 = \left(\frac{1}{\sqrt{2\pi} \Delta x}\right)^{d} \exp\left[-\frac{\bold{x}^2}{2(\Delta x)^2}\right] \, .
\end{eqnarray}
This allows us to compute the integral (\ref{<Delta_l>-1}) exactly, yielding 
\begin{eqnarray} \label{<Delta_l>-2}
\braket{\Delta l}_{\psi} \propto \sigma_{\psi}(\Delta t) = \frac{\hbar}{2m}\frac{\Delta t}{\Delta x} \sqrt{1 + \frac{4m^2}{\hbar^2}\frac{(\Delta x)^4}{(\Delta t)^2}} \, . 
\end{eqnarray}
where $ \sigma_{\psi}(\Delta t)$ denotes the standard deviation of the probability distribution $|\psi_{\Delta t}(\bold{x})|^2$, given the initial Gaussian distribution (\ref{psi_0}). 
However, even for non-Gaussian initial profiles, the quantitive results are similar, differing only to within numerical factors of order unity \cite{Abbott:1979bh}. 

For this reason, the definition of $\braket{\Delta l}_{\psi}$ (\ref{<Delta_l>-1}) is somewhat arbitrary and we may instead use 
\begin{eqnarray} \label{<Delta_l>-3}
\braket{\Delta l}_{\psi} &=& \braket{\psi | \hat{U}^{\dagger}(\Delta t) \hat{\bold{x}}^2 \hat{U}(\Delta t) | \psi}^{1/2} 
\nonumber\\
&=& \left(\int \bold{x}^2 |\psi_{\Delta t}(\bold{x})|^2 {\rm d}^dx\right)^{1/2} \, ,
\end{eqnarray}
as also noted in \cite{Abbott:1979bh}. 
Since each measurement effectively resets the coordinate origin to the centre of the newly resolved probability distribution, which is equivalent to $|\psi_{0}(\bold{x})|^2$ (\ref{psi_0}) in the new coordinates, using the definition (\ref{<Delta_l>-3}) in place of (\ref{<Delta_l>-1}) gives $\braket{\Delta l}_{\psi} = \Delta_{\psi}x$. 
Here, we use $\Delta_{\psi}x$ to denote the standard deviation of a general wave function, which is not necessarily a Gaussian. 

For Gaussian distributions, Eq. (\ref{<Delta_l>-3}) gives $\braket{\Delta l}_{\psi} = \sigma_{\psi}(\Delta t)$, resulting in exact equality, rather than just proportionality, in Eq. (\ref{<Delta_l>-2}). 
After $N$ measurements, corresponding to a total time interval $T = N\Delta t$, the total distance travelled by the particle is then
\begin{eqnarray} \label{<l>-1}
\braket{l}_{\psi} = N\braket{\Delta l}_{\psi} = N\sigma_{\psi}(\Delta t) \, .
\end{eqnarray}

Finally, combining $\Delta E \simeq (\Delta p)^2/(2m)$, where $\Delta p = \hbar/(2\Delta x)$ is the width of the initial momentum space Gaussian $|\tilde{\psi}_{0}(\bold{p})|^2$, with the uncertainty relation for energy and time, $\Delta E\Delta t \gtrsim \hbar/2$, gives
\begin{eqnarray} \label{Delta t}
\Delta t \gtrsim \frac{4m(\Delta x)^2}{\hbar} \, . 
\end{eqnarray}
Substituting (\ref{Delta t}) into (\ref{<Delta_l>-2}) and (\ref{<Delta_l>-2}) into (\ref{<l>-1}) yields
\begin{eqnarray} \label{<l>-2}
\braket{l}_{\psi} \simeq \frac{N\hbar}{2m}\frac{\Delta t}{\Delta x} \, .  
\end{eqnarray}
This shows that the path of the particle is self-similar, i.e., fractal, since changing the resolution such that $\Delta x \rightarrow \Delta x' = \gamma \Delta x$ simply rescales the observed length of the path by a factor of $\gamma^{-1}$ \cite{Falconer,Abbott:1979bh}. 

By analogy with Eq. (\ref{L_H}), Abbott and Wise defined the Hausdorff length of a QM particle path as \cite{Abbott:1979bh}
\begin{eqnarray} \label{<L_H>-1}
\braket{L_{\rm H}}_{\psi} = \braket{l}_{\psi} \, . \, (\Delta x)^{D_{\rm H}-1} \, . 
\end{eqnarray}
Substituting from (\ref{<l>-2}) into (\ref{<L_H>-1}) it is easy to see that $\braket{L_{\rm H}}_{\psi}$ is independent of $\Delta x$ if $D_{\rm H} = 2$, giving
\begin{eqnarray} \label{<L_H>-2}
\braket{L_{\rm H}}_{\psi} \propto N\frac{\hbar\Delta t}{2m} \, . 
\end{eqnarray}

Let us now switch to the momentum space picture and define the momentum space path length as
\begin{eqnarray} \label{<tilde_l>-1}
\braket{\tilde{l}}_{\psi} = N\braket{\Delta \tilde{l}}_{\psi} = N\tilde{\sigma}_{\psi}(\Delta t) \, ,
\end{eqnarray}
where $\tilde{\sigma}_{\psi}(\Delta t)$ is the width of the momentum space Gaussian, $|\tilde{\psi}_{\Delta t}(\bold{p})|^2$, and $\tilde{\sigma}_{\psi}(0) = \Delta p = \hbar/(2\Delta x)$. 
In the limit $\Delta t \gtrsim \hbar m/(\Delta p)^2$, which is equivalent to the condition (\ref{Delta t}), we then have
\begin{eqnarray} \label{tilde_sigma}
\tilde{\sigma}_{\psi}(\Delta t) = \frac{\hbar}{2\sigma_{\psi}(\Delta t)} \simeq \frac{m\Delta x}{\Delta t} = \frac{\hbar}{2}\frac{m}{\Delta p\Delta t} \, . 
\end{eqnarray} 
Defining the momentum space Hausdorff length as 
\begin{eqnarray} \label{<tilde_L_H>}
\braket{\tilde{L}_{\rm H}}_{\psi} = \braket{\tilde{l}}_{\psi} \, . \, (\Delta p)^{\tilde{D}_{\rm H}-1} \, ,
\end{eqnarray}
we see that $\tilde{D}_{\rm H} = 2$, since $\braket{\tilde{l}}_{\psi} \propto (\Delta p)^{-1}$. 
Hence, in canonical QM, the Hausdorff dimensions of the particle path in both the position and momentum space representations are equal, at least for relatively large time-intervals, $\Delta t \gtrsim 4m(\Delta x)^2/\hbar = \hbar m/(\Delta p)^2$, for which $D_{\rm H} = \tilde{D}_{\rm H} = 2$. 

Note that these results also hold if we choose to resolve the path of the particle in momentum space directly, by performing successive momentum measurements, rather than measurements of position.
In fact, in the experimental scenario considered here, a position measurement with finite accuracy $\sim (\Delta x)^d$ constitutes a de facto measurement of momentum, with accuracy $\sim (\Delta p)^d$, where $\Delta p = \hbar/(2\Delta x)$, and vice versa.  

Before concluding this section, we now return to the position space representation, in order to perform a more careful analysis of the condition $d\braket{L_{\rm H}}_{\psi}/d(\Delta x) = 0$ (*). 
As we will now show, this allows us to derive an important result that was not presented in Abbott and Wise's original analysis \cite{Abbott:1979bh}. 

Treating $\Delta x$ and $\Delta t$ as independent variables, and combining (\ref{<Delta_l>-2}), (\ref{<l>-1}) and (\ref{<L_H>-1}) with (*), gives rise to the polynomial equation
\begin{eqnarray} \label{polynomial-1}
(\Delta x)^{D_{\rm H}-2} + \frac{(D_{\rm H}-2)}{D_{\rm H}} \frac{\hbar^2}{4m} (\Delta t)^2 (\Delta x)^{D_{\rm H}-6} = 0 \, .
\end{eqnarray}
Clearly, this equation can be solved by allowing $\Delta t \rightarrow \infty$ and setting $D_{\rm H} = 2$. 
However, it may also be solved another way, by parameterising $\Delta t$ in terms of $(\Delta x)^2$ as  
\begin{eqnarray} \label{Delta_t_param}
\Delta t = \xi \, . \, \frac{4m(\Delta x)^2}{\hbar} \, , 
\end{eqnarray}
where $\xi \gtrsim 1$. 
Substituting into (\ref{polynomial-1}) then gives
\begin{eqnarray} \label{D_H}
D_{\rm H}(\xi) = \frac{2}{1+\frac{1}{4\xi^2}} \, , 
\end{eqnarray}
together with
\begin{eqnarray} \label{<L_H>-2*}
\braket{L_{\rm H}}_{\psi} \propto 2N \left(1+\frac{1}{4\xi^2}\right)^{1/2} \xi^{-\left(\frac{D_{\rm H}(\xi)-2}{2}\right)}\sqrt{\frac{\hbar\Delta t}{4m}}^{D_{\rm H}(\xi)} \, .
\end{eqnarray}
These results follow from imposing $d\braket{L_{\rm H}}_{\psi}/d(\Delta x)|_{\Delta x = \sqrt{\hbar\Delta t/4m\xi}} = 0$, rather than simply $d\braket{L_{\rm H}}_{\psi}/d(\Delta x) = 0$, but the former is the most physically relevant condition for the experimental scheme considered in \cite{Abbott:1979bh}.  

This analysis reflects the fact that, although the resolution $\Delta x$ and time-interval $\Delta t$ can be chosen independently by the experimenter tracking the path of the particle, any value of $\Delta t$ must, necessarily, be some multiple of the minimum value $4m(\Delta x)^2/\hbar$ (\ref{Delta t}). 
The result (\ref{D_H}) suggests that which multiple is chosen affects the fractal properties of the path, with shorter sampling times yielding lower values of the Hausdorff dimension. 
This has a clear physical interpretation and reflects the fact that sampling the path disrupts the process of free quantum diffusion. 

The quantum diffusion is akin to Brownian motion \cite{Nelson_1966} which, over time, builds up the self-similar fractal path of the particle. 
To build a totally self-similar path, i.e., one which is self-similar on all scales from $\Delta x \rightarrow 0$ up to $\Delta x \rightarrow \infty$, requires infinite time. 
If left undisturbed, the quantum motion of the particle will eventually cause its path to cover a two-dimensional hypersurface, embedded as a fractal within the $d$-dimensional background space.  
Only for $\Delta t \rightarrow \infty$ is the free-particle path fractal complete, giving $D_{\rm H} = 2$. 

In the real world process of sampling the path, each measurement effectively resets the diffusion process back to its initial conditions. 
The `true' late-time fractal path remains only partially constructed at any finite time-interval, which manifests as a decrease in the Hausdorff dimension. 
This is the physical meaning of Eq. (\ref{D_H}). 

It is important to realise that, by sampling the path with nonzero $\Delta x$ at finite $\Delta t$, we are not sampling a perfect fractal, i.e., an infinitely self-similar path, with finite resolution. 
Instead, we are sampling an imperfect fractal that has not had time to develop self-similarity on all scales. 
Nonetheless, when $\Delta t$ is several orders of magnitude larger than $4m(\Delta x)^2/\hbar$, the particle has had sufficient time to generate self-similarity on scales up to $\sim (\Delta x)^d$, yielding the `correct' measured value of the path-fractal Hausdorff dimension. 
In this way, the QM path can be thought of as a fractal created `in reverse', being built up from self-similar patterns on the smallest scales, over the smallest time-intervals, and achieving self-similarity up to very large scales only at late times.  

Hence, from Eq. (\ref{D_H}) it is clear that the result obtained by Abbott and Wise, $D_{\rm H} = 2$, corresponds to the large time-interval limit, $\xi \rightarrow \infty$ ($\Delta t \rightarrow \infty$). 
In this regime, the general expression for the Hausdorff length, Eq. (\ref{<L_H>-2*}), reduces to Eq. (\ref{<L_H>-2}). 
However, as $\Delta t$ approaches its smallest permissible value, i.e., for $\xi \rightarrow 1$ ($\Delta t = 4m(\Delta x)^2/\hbar$), we obtain $D_{\rm H} = 8/5 = 1.6$. 
As the parameter $\xi$ varies in the range $1 \leq \xi < \infty$, the Hausdorff dimension of the path varies in the range $8/5 \leq D_{\rm H} < 2$, according to (\ref{D_H}). 
We stress that, strictly, the result obtained in \cite{Abbott:1979bh} corresponds only to the asymptotic limit, $\Delta t \rightarrow \infty$, and that the Hausdorff dimension of the particle path in real space may drop somewhat below $D_{\rm H} = 2$, if the time-interval between successive measurements is short enough.  

In the momentum space picture we use Eqs. (\ref{<Delta_l>-2}), (\ref{<tilde_L_H>}) and (\ref{Delta_t_param}), together with the fact that $\tilde{\sigma}_{\psi}(\Delta t) = (\hbar/2)\sigma^{-1}_{\psi}(\Delta t)$ and $\tilde{\sigma}_{\psi}(0) = \Delta p = \hbar/(2\Delta x)$, to impose the condition $d\braket{\tilde{L}_{\rm H}}_{\psi}/d(\Delta p) = 0$ (**). 
This leads to the polynomial equation
\begin{eqnarray} \label{polynomial-2}
(\Delta p)^{\tilde{D}_{\rm H}-2} + \frac{4(\tilde{D}_{\rm H}-2)}{\tilde{D}_{\rm H}} \frac{(\Delta t)^2}{\hbar^2 m^2} (\Delta p)^{\tilde{D}_{\rm H}+2} = 0 \, .
\end{eqnarray}
Substituting $\Delta t = \xi \, . \, m\hbar/(\Delta p)^2$ then gives
\begin{eqnarray} \label{tilde_D_H}
\tilde{D}_{\rm H}(\xi) = \frac{2}{1+\frac{1}{4\xi^2}} \, ,
\end{eqnarray}
which is equivalent to imposing the condition $d\braket{\tilde{L}_{\rm H}}_{\psi}/d(\Delta p)|_{\Delta p = \sqrt{\xi m\hbar/\Delta t}} = 0$. 

From Eqs. (\ref{polynomial-1}) and (\ref{polynomial-2}) we see that, for all values of the parameter $\xi$, the Hausdorff dimensions of the particle path in the position and momentum space representations are equal, $D_{\rm H} = \tilde{D}_{\rm H}$, and vary within the range $[8/5,2)$. 
The upper limit corresponds to the result obtained by Abbott and Wise but the lower limit, which is valid for the shortest possible time-intervals, was previously unknown. 
In the following sections, we will show how the presence of GUP- and EUP-induced corrections alters these results.  

\subsection{Previous GUP models} \label{Sec.2.2} 

In \cite{Nicolini:2010dj,NiednerThesis}, Nicolini and Niedner analysed a GUP model based on the modified momentum space volume
\begin{eqnarray} \label{d_tilde_V}
{\rm d}\tilde{V} = \exp\left[-\frac{\bold{P}^2}{2(\hbar/l)^2}\right] {\rm d}^{d}P \, , 
\end{eqnarray}
where $l$ is the minimum length. 
To first order in the expansion of the Gaussian, this corresponds to the modified commutator 
\begin{eqnarray} \label{mod_comm-1}
[\hat{X}^i,\hat{P}_j] = i\hbar \, \delta^{i}{}_{j}\left(1 + \frac{\hat{\bold{P}}^2}{2(\hbar/l)^2}\right) \hat{\openone}
\end{eqnarray}
where 
\begin{eqnarray} \label{P_j}
\hat{P}_j = \int P_j \proj{\bold{P}}\exp\left[-\frac{\bold{P}^2}{2(\hbar/l)^2}\right] {\rm d}^{d}P \, 
\end{eqnarray}
is the modified momentum-measurement operator and 
\begin{eqnarray} \label{}
&&\braket{\bold{P}|\bold{P}'} = \exp\left[\frac{\bold{P}^2}{2(\hbar/l)^2}\right]\delta^{d}(\bold{P}-\bold{P}') \, ,
\nonumber\\
&&\int \proj{\bold{P}}\exp\left[-\frac{\bold{P}^2}{2(\hbar/l)^2}\right] {\rm d}^{d}P =  \hat{\openone} \, . 
\end{eqnarray}
Throughout the rest of this paper we use capital letters to refer to modified position and momentum operators that give rise to GURs and lower case letters to refer to their canonical QM counterparts. 
Thus, in this section, $\hat{X}^i$ and $\hat{P}_j$ denote the position and momentum operators of the modified commutator model (\ref{mod_comm-1}), whereas, in Sec. \ref{Sec.4}, the same symbols are used to denote the modified smeared space operators derived in \cite{Lake:2018zeg,Lake:2019nmn,Lake2020-2}. 
Clearly, when the minimum length is set equal to the $(3+1)$-dimensional Planck length, $l \equiv l_{\rm Pl} = \sqrt{\hbar G/c^3}$, Eq. (\ref{mod_comm-1}) yields the GUP (\ref{GUP}).

In this model, the Hausdorff dimension is determined by following steps analogous to those outlined in Sec. \ref{Sec.1}, substituting the modified momentum-measurement operator (\ref{P_j}) in place the the canonical operator $\hat{p}_j = \int p_j \proj{\bold{p}^2}{\rm d}^{d}p$. 
This gives rise to a modified free-particle Hamiltonian and, hence, to the modified time-evolution operator
\begin{eqnarray} \label{}
\hat{\mathcal{U}}(t) = \exp\left[-\frac{\hat{\bold{P}}^2}{2m}t \right] \, ,  
\end{eqnarray}
where the components of $\hat{\bold{P}}$ are obtained from Eq. (\ref{P_j}). 
The path length traversed in time $\Delta t$ may then be defined as
\begin{eqnarray} \label{<Delta_l>-1*}
\braket{\Delta L}_{\psi} = \braket{\psi | \hat{\mathcal{U}}^{\dagger}(\Delta t) |\hat{\bold{X}}| \hat{\mathcal{U}}(\Delta t) | \psi} \, ,
\end{eqnarray}
or 
\begin{eqnarray} \label{<Delta_l>-3*}
\braket{\Delta L}_{\psi} = \braket{\psi | \hat{\mathcal{U}}^{\dagger}(\Delta t) \hat{\bold{X}}^2 \hat{\mathcal{U}}(\Delta t) |\psi}^{1/2} \, ,
\end{eqnarray}
by analogy with the canonical theory. 
Here, we choose the definition (\ref{<Delta_l>-3*}), for convenience. 
Again, since each measurement effectively resets the coordinate origin to the centre of the newly resolved wave function, adopting (\ref{<Delta_l>-3*}) gives
\begin{eqnarray} \label{<Delta_l>-2*}
\braket{\Delta L}_{\psi} = \Delta_{\psi}X \, . 
\end{eqnarray}

It may then be shown that, for an initially Gaussian wave packet with standard deviation $\Delta_{\psi} X(0) \equiv \sigma_{\psi}(0) = \Delta X$, the total path length traversed in $T = N\Delta t$ is \cite{Nicolini:2010dj,NiednerThesis}
\begin{eqnarray} \label{<L>-1}
\braket{L}_{\psi} &=& \frac{N\hbar}{m}\frac{\Delta t}{\Delta X}\left(1 + \frac{l^2}{(\Delta X)^2}\right)^{-\frac{d+1}{2}}
\nonumber\\
&\times& \sqrt{1 + \left(1 + \frac{l^2}{(\Delta X)^2}\right)^2\frac{4m^2(\Delta X)^4}{\hbar^2(\Delta t)^2}} \, . 
\end{eqnarray}
For $\Delta t \gtrsim 4m(\Delta X)^2/\hbar$ and $\Delta X \gtrsim l$, this gives
\begin{eqnarray} \label{<L_H>-1*}
\braket{L_{\rm H}}_{\psi} \propto (\Delta X)^{D_{\rm H}-2}\left(1 + \frac{l^2}{(\Delta X)^2}\right)^{-\frac{d+1}{2}} \, . 
\end{eqnarray}

Nicolini and Niedner then claimed that imposing $d\braket{L_{\rm H}}_{\psi}/d(\Delta X) = 0$ yields \cite{Nicolini:2010dj,NiednerThesis}
\begin{eqnarray} \label{D_H_Nicolini-1}
D_{\rm H} = 2 - \frac{d+1}{1 + (\Delta X)^2/l^2} \, . 
\end{eqnarray} 
Next, they showed that the spectral dimension of a probability density obeying the classical diffusion equation, with diffusion coefficient $s$, and in the presence of a minimum length $l$ in $d$ spatial dimensions, is
\begin{eqnarray} \label{spectral_dim}
\mathbb{D} = \frac{s}{s + l^2} \, D \, , 
\end{eqnarray}
where $D = d+1$. 
Taking $|\psi_{\Delta t}|^2(\bold{X}) = |\braket{\bold{X}| \hat{\mathcal{U}}(\Delta t)|\psi}|^2$ as the probability density, Wick rotating the diffusion equation to obtain the canonical Schr{\" o}dinger equation, and identifying the diffusion coefficient with $(\Delta X)^2$ then gives
\begin{eqnarray} \label{D_H_Nicolini-2}
D_{\rm H} = 2 - (D-\mathbb{D}) \, . 
\end{eqnarray}

This is a very nice result, which neatly connects the Hausdorff and spectral dimensions of the path of a non-relativistic particle with the topological dimension of the spacetime it propagates in. 
However, its validity depends on the validity of Eq. (\ref{D_H_Nicolini-1}). 
It is straightforward to show that this formula is derived by treating $D_{\rm H}$ as a constant when taking the derivate of $\braket{L_{\rm H}}_{\psi}$ with respect to $\Delta X$. 
This leads to a contradiction, since $D_{\rm H}$ in Eq. (\ref{D_H_Nicolini-1}) is a function of $\Delta X$. 
Indeed, by inspection, it is clear that the expression for $\braket{L_{\rm H}}_{\psi}$ given in Eq. (\ref{<L_H>-1*}) satisfies the condition $d\braket{L_{\rm H}}_{\psi}/d(\Delta x) = 0$ only in the limit $\Delta X \gg l$. 
In this regime, the particle path is too coarse grained for the GUP-induced corrections to alter its observable characteristics, and the Hausdorff dimension is $D_{\rm H}=2$, as in canonical QM. 

Nonetheless, assuming $D_{\rm H} =  D_{\rm H}(\Delta X)$, it is also straightforward to show that Eq. (\ref{D_H_Nicolini-1}) satisfies the condition $d\braket{L_{\rm H}}_{\psi}/d(\Delta X) = 0$, at least approximately for $\Delta X \simeq l$, since ${\rm ln}(\Delta X) \, . \, d D_{\rm H}/d(\Delta X) \simeq 0$ in this regime. 
Therefore, taking the limits $\Delta X = l$ and $\Delta X \gg l$, separately, this analysis still suggests that the Hausdorff dimension of the particle path in the GUP model (\ref{mod_comm-1}) varies within the range $D_{\rm H} \in [2-(d+1)/2,2)$, as claimed in \cite{Nicolini:2010dj,NiednerThesis}. 
However, it is important to note that Eq. (\ref{D_H_Nicolini-2}) can be derived from the condition $d\braket{L_{\rm H}}_{\psi}/d(\Delta x) = 0$ only when the spatial resolution is set close to the $D$-dimensional Planck scale, $\Delta X \simeq l \simeq l^{(D)}_{\rm Pl}$.  

Finally, we must also account for the fact that the self-similar path takes time to develop by parameterising the time-interval as
\begin{eqnarray} \label{xi}
\Delta t = \xi \, . \, \frac{4m(\Delta X)^2}{\hbar} \, ,
\end{eqnarray}
where $\xi \gtrsim 1$, as in the canonical theory. 
Similarly, we can parameterise the minimum length in terms of the width of the detecting apparatus, such that
\begin{eqnarray} \label{epsilon}
l = \epsilon \, . \, \Delta X \, , 
\end{eqnarray}
where $\epsilon \lesssim 1$. 
To analyse the smallest possible resolutions, $\Delta X \simeq l$, over the smallest possible time-scales, $\Delta t \simeq 4m l^2/\hbar$, we must first treat $\Delta X$, $\Delta t$ and $l$ as independent variables when applying $d\braket{L_{\rm H}}_{\psi}/d(\Delta X) = 0$. 
We then impose the parameterisations (\ref{xi}) and (\ref{epsilon}), which is equivalent to applying the condition $d\braket{L_{\rm H}}_{\psi}/d(\Delta X)|_{\Delta X = \sqrt{\hbar \Delta t/4m\xi} = l/\epsilon} = 0$. 

The first step leads to the polynomial equation
\begin{eqnarray} \label{polynomial-3}
&&(\Delta X)^{D_{\rm H}-2} + (D_{\rm H}-2) \frac{\hbar^2(\Delta t)^2}{8m^2} \frac{(\Delta X)^{D_{\rm H}-4}}{[(\Delta X)^2+l^2]}
\nonumber\\
&+& \frac{(D_{\rm H}-2)}{2}[(\Delta X)^2+l^2](\Delta X)^{D_{\rm H}-4} 
\nonumber\\
&+&
\frac{(d+1)l^2}{[(\Delta X)^2+l^2]^2}  \frac{\hbar^2(\Delta t)^2}{8m^2} (\Delta X)^{D_{\rm H}-4} 
\nonumber\\
&\times& \left[1 + \frac{4m^2}{\hbar^2}\frac{[(\Delta X)^2+l^2]^2}{(\Delta t)^2}\right] \, , 
\end{eqnarray}
which reduces to Eq. (\ref{polynomial-1}) when $l \rightarrow 0$. 
Substituting from Eqs. (\ref{xi})-(\ref{epsilon}) then gives
\begin{eqnarray} \label{D_H_GUP_NN}
&&D_{\rm H}(\xi,\epsilon) = 2 -
\nonumber\\
&&(d+1) \frac{\epsilon^2}{1+\epsilon^2}\left[1 + \frac{2}{4\xi^2 + (1+\epsilon^2)^2} . \frac{(1+\epsilon^2)^2}{(d+1)\epsilon^2}\right]. 
\end{eqnarray}

For $\epsilon \rightarrow 0$, Eq. (\ref{D_H_GUP_NN}) reduces to Eq. (\ref{D_H}), as required, but taking the limit $\xi \rightarrow \infty$ with $\epsilon > 0$ gives 
\begin{eqnarray} \label{fixed}
D_{\rm H}(\epsilon) = 2 - (d+1) \frac{\epsilon^2}{1+\epsilon^2} \, . 
\end{eqnarray}
This is equivalent to Eq. (\ref{D_H_Nicolini-1}), expressed in terms of our new parameter $\epsilon = l/\Delta X$ (\ref{epsilon}). 
However, here, there is no contradiction, since Eq. (\ref{fixed}) is derived from the condition $d\braket{L_{\rm H}}_{\psi}/d(\Delta X)|_{\Delta X = \sqrt{\hbar \Delta t/4m\xi} = l/\epsilon} = 0$ rather than simply $d\braket{L_{\rm H}}_{\psi}/d(\Delta X) = 0$. 

The former is the most physically relevant condition, for the experimental scenario considered, and our analysis reflects the fact that $\Delta t$ and $\Delta X$ may be chosen independently, as in canonical QM, as well as the fact that both are independent of the minimum length $l$. 
It also accounts for the fact that the chosen value of $\Delta t$ must be a multiple of the minimum possible time-interval, $(\Delta t)_{\rm min} = 4m(\Delta X)^2/\hbar$ (\ref{xi}), and that the chosen value of $\Delta X$ must be a multiple of $l$. 
Equation (\ref{fixed}) may then be combined with Eq. (\ref{spectral_dim}) to give Eq. (\ref{D_H_Nicolini-2}), which is thereby shown to be valid on all scales, without introducing contradictory assumptions. 
 
Finally, we note that, for $\epsilon \rightarrow 1$, Eq. (\ref{fixed}) becomes $D_{\rm H} = 2-(d+1)/2$. 
Imposing $D_{\rm H} > 1$, i.e., requiring that the Hausodorff dimension of the fractal strictly exceed the topological dimension of a classical particle path \cite{Falconer}, then yields $d < 1$. 
This suggests that the fractal properties of the particle path can only be probed on the smallest scales in at most one spatial dimension. 
In higher-dimensional spaces, the path of the particle does not exhibit fractal properties at the minimum length scale. 
Similarly, imposing $D_{\rm H} > 0$ requires $d < 3$. 

How can we interpret this result? 
In \cite{Nicolini:2010dj}, it was proposed that $D_{\rm H} \rightarrow 0$ corresponds to the trans-Plankian regime in which the path of the particle completely disintegrates due to Planck-scale fluctuations induced by the GUP. 
This is physically reasonable since negative values of the Hausdorff dimension correspond to empty sets \cite{Falconer}. 
Therefore, we combine the condition $D_{\rm H} \geq 0$ with Eq. (\ref{fixed}), which implies that $\epsilon^2 \leq 2/(D+2)$ or, equivalently, $\Delta X \geq \sqrt{(D+2)/2} \, l$. 

In other words, in the GUP model based on the modified phase space volume (\ref{d_tilde_V}) there exists a fundamental limit to the scale at which a particle path can be meaningfully resolved (as expected). 
For $d > 1$, this is somewhat above the actual minimum volume, $l^d$, and its exact value is determined by the dimensionality of the background, such that $\Delta V_{\rm min} \simeq (\sqrt{(D+2)/2} \, l)^{D-1}$. 
Hence, although the analysis presented here differs somewhat from that given by Nicolini and Niedner \cite{Nicolini:2010dj,NiednerThesis}, we validate their claim that, in the GUP model (\ref{mod_comm-1}), the fractal properties of the particle path depend on the dimensions of the spacetime, $D$. 

\section{Recap of the smeared space model} \label{Sec.3}

In this section, we review the basic formalism of the smeared space model, originally presented in \cite{Lake:2018zeg,Lake:2019nmn,Lake2020-2}. 
We then highlight important differences between smeared space and previous GUR models based on modified commutation relations. 


\subsection{Basic formalism} \label{Sec.3.1} 

In \cite{Lake:2018zeg}, a new model of quantum geometry was proposed in which each point ${\bf x}$ in the classical background is associated with a vector in a Hilbert space,   
\begin{eqnarray} \label{g_x}
\ket{g_{\bold{x}}} = \int g(\bold{x}'-\bold{x}) \ket{\bold{x}'} {\rm d}^{d}{\rm x}' \, ,
\end{eqnarray}
where $\braket{g_{\bold{x}}|g_{\bold{x}}}=1$. 
This is used to describe a form of nonlocal geometry that is intrinsically quantum in nature, so that the width of $|g(\bold{x}'-\bold{x})|^2$ is assumed to be of the order of the Planck length \cite{Lake:2018zeg,Lake:2019nmn,Lake2020-2}. 

It is well-known that classical nonlocal geometries, such as (\ref{d_tilde_V}), can be generated by first identifying each point in the classical manifold with a Dirac delta function, $\delta^d(\bold{x - x'})$. 
Nonlocality is then introduced by smearing each delta into a finite-width probability distribution $P({\bf x-x'})$. 
(For example, a normalised Gaussian in the model considered by Nicolini and Niedner \cite{Nicolini:2010dj,NiednerThesis}.)
In this case, no new degrees of freedom are introduced, beyond those present in canonical QM, since $\bold{x}'$ is simply a parameter that determines the position of $P$. 

The smeared space model introduced in \cite{Lake:2018zeg,Lake:2019nmn,Lake2020-2} is different in that it first associates each point ${\bf x'}$ with a rigged basis vector of a Hilbert space, $\ket{\bold{x'}}$. 
The latter is then smeared to produce the normalised state (\ref{g_x}). 
In this case, $\braket{\bold{x'} | g_{\bold{x}}} = g(\bold{x}'-\bold{x})$ is a genuine quantum mechanical amplitude not a probability distribution. 
It has dimensions of ${\rm (length)}^{-d/2}$ not ${\rm (length)}^{-d}$ and, in principle, can possess nontrivial phase information. 
In this model, $\ket{g_{\bold{x}}}$ represents the state of a Planck-scale localised `point' in the quantum geometry. 
Each point in the classical geometry is then smeared into a Planck-volume localised superposition of {\it all} points in the background space by imposing the map
\begin{eqnarray} \label{smearing_map}
S: \ket{\bf{x}} \mapsto \ket{\bf{x}} \otimes \ket{g_{\bf{x}}} \, .
\end{eqnarray}

The smearing map (\ref{smearing_map}) may be visualised as follows: for each point $\bold{x} \in \mathbb{R}^d$ in the classical geometry it generates one whole `copy' of $\mathbb{R}^d$, thereby doubling the size of the classical phase space. 
The resulting smeared geometry is represented by a $2d$-dimensional volume in which each point $(\bf{x},\bf{x}')$ is associated with a quantum probability amplitude, $g(\bold{x}'-\bold{x})$. 
This is interpreted as the amplitude for the transition $\bold{x} \leftrightarrow \bold{x}'$ and the higher-dimensional space is interpreted as a superposition of $d$-dimensional geometries \cite{Lake:2018zeg,Lake:2019nmn,Lake2020-2}. 

In the nonrelativisitc limit, each geometry in the smeared superposition of geometries is Euclidean, but differs from all others by the pair-wise exchange of at least two points \cite{Lake:2019nmn,Lake2020-2}. 
It is assumed that the interchange $\bold{x} \leftrightarrow \bold{x}'$ exchanges the canonical amplitudes, $\psi(\bold{x}) \leftrightarrow \psi(\bold{x}')$, which leads to additional fluctuations in the observed position of the particle, over and above those obtained in canonical QM. 
We now review, briefly, how these fluctuations give rise to GURs. 

For simplicity, we may take $|g(\bold{x}'-\bold{x})|^2$ to be a normalised Gaussian centred on $\bf{x}' = \bf{x}$, but, here, $\bf{x}'$ is no longer just a parameter. 
By introducing the tensor product structure (\ref{smearing_map}) we have doubled the number of degrees of freedom of the theory, vis-{\` a}-vis canonical QM. 
Those in the left-hand subspace, labelled by $\bold{x}$, represent the degrees of freedom of a canonical quantum particle, whereas those in the right-hand subspace, labelled by $\bold{x'}$, determine the influence of fluctuations in the background geometry. 
The action of $S$ on $\ket{\bold{x}}$ (\ref{smearing_map}) then induces a map on the canonical quantum state, $\ket{\psi} = \int \psi(\bold{x})\ket{\bold{x}} {\rm d}^{3}{\rm x}$, such that
\begin{eqnarray} \label{psi->Psi}
S : \ket{\psi} \mapsto \ket{\Psi} \, , 
\end{eqnarray}
where
\begin{eqnarray} \label{|Psi>_position_space}
\ket{\Psi} = \int\int \psi(\bold{x}) g(\bold{x}{'}-\bold{x}) \ket{\bold{x},\bold{x}{\, '}} {\rm d}^{d}{\rm x}{\rm d}^{d}{\rm x}' \, .
\end{eqnarray} 

The square of the smeared-state wave function, $|\Psi(\bold{x},\bold{x}')|^2 = |\psi(\bold{x})|^2|g(\bold{x}'-\bold{x})|^2$, represents the probability distribution associated with a quantum particle propagating in a quantum superposition of geometries \cite{Lake:2018zeg}. 
Because $|\psi(\bold{x})|^2$ represents the probability of finding the particle at the fixed classical point $\bf{x}$ in canonical QM, $|\psi(\bold{x})|^2|g(\bold{x}'-\bold{x})|^2$ represents the probability that it will now be found, instead, at a new point $\bold{x}'$. 
If $g(\bold{x})$ is a Gaussian centred on the origin, $\bold{x}'=\bold{x}$ remains the most likely value, but fluctuations within a volume of order $\sim \sigma_g^{d}$, where $\sigma_g$ is the standard deviation of $|g(\bold{x})|^2$, remain relatively likely \cite{Lake:2018zeg,Lake:2019nmn,Lake2020-2}. 
Furthermore, since an observed position `$\bold{x}'$' cannot determine which point(s) underwent the transition $\bold{x} \leftrightarrow \bold{x}'$ in the smeared superposition of geometries, we must sum over all possibilities by integrating the joint probability distribution $|\Psi(\bold{x},\bold{x}')|^2$ over ${\rm d}^d{\rm x}$, yielding
\begin{eqnarray} \label{EQ_XPRIMEDENSITY}
\frac{{\rm d}^{d}P(\bold{x}' | \Psi)}{{\rm d}{\rm x}'^{d}} = \int |\Psi(\bold{x},\bold{x}')|^2 {\rm d}^d{\rm x} = (|\psi|^2 * |g|^2)(\bold{x}') \, ,
\end{eqnarray}
where the star denotes a convolution. 
In this formalism, only primed degrees of freedom represent measurable quantities, whereas unprimed degrees of freedom are physically inaccessible \cite{Lake:2018zeg,Lake:2019nmn,Lake2020-2}.

The variance of a convolution is equal to the sum of the variances of the individual functions, so that the probability distribution (\ref{EQ_XPRIMEDENSITY}) gives rise to the GUR 
\begin{eqnarray} \label{X_uncertainty}
(\Delta_\Psi X^{i})^2 = (\Delta_\psi x'^{i})^2 + (\Delta_gx'^i)^2 \, .
\end{eqnarray}
It is straightforward to verify that (\ref{X_uncertainty}) is obtained from the standard braket construction $(\Delta_\Psi X^{i})^2 = \braket{\Psi |(\hat{X}^{i})^{2}|\Psi} - \braket{\Psi|\hat{X}^{i}|\Psi}^2$, where  
\begin{eqnarray} \label{X_operator}
\hat{X}^{i} = \int x'^{i} \, {\rm d}^{d} \hat{\mathcal{P}}_{\bold{x}'} = \hat{\openone} \otimes \hat{x}'^{i} 
\end{eqnarray}
is the generalised position-measurement operator and ${\rm d}^{d}\hat{\mathcal{P}}_{\bold{x}'} = \hat{\openone} \otimes \ket{\bold{x}'}\bra{\bold{x}'}{\rm d}^{d}{\rm x}'$.

Next, we note that the HUP, expressed here in terms of the physically accessible primed variables,
\begin{eqnarray} \label{HUP}
\Delta_{\psi} x'^{i} \Delta_{\psi} p'_{j} \geq \frac{\hbar}{2} \delta^{i}{}_{j} \, , 
\end{eqnarray}
holds independently of Eq. (\ref{X_uncertainty}). 
Combining the two and identifying the standard deviation of $|g({\bf x})|^2$ with the $D$-dimensional Planck length such that
\begin{equation} \label{sigma_g} 
\Delta_{g}x'^{i} = \sigma_g^{i} = 2^{\frac{1}{D-2}}l^{(D)}_{\rm Pl} \, ,  
\end{equation}
then yields
\begin{eqnarray} \label{smeared_GUP}
\Delta_\Psi X^{i} \gtrsim \frac{\hbar}{2\Delta_{\psi} p'_{j}}\delta^{i}{}_{j}\left[1 + \alpha(\Delta_{\psi} p'_{j})^2\right] \, , 
\end{eqnarray}
where $\alpha = 4(m^{(D)}_{\rm Pl}c)^{-2}$, to first order in the expansion \cite{Lake:2018zeg}. 
For $\Delta_{\psi} x'^{i} \gg \sigma_g^{i} \simeq l^{(D)}_{\rm Pl}$, we have $\Delta_\Psi X^{i} \simeq \Delta_{\psi} x'^{i}$, so that, in this limit, Eq. (\ref{smeared_GUP}) reduces to the GUP (\ref{GUP}) when $D = 3+1$. 

In the momentum space picture, the composite matter-plus-geometry state $\ket{\Psi}$ is expanded as 
\begin{eqnarray} \label{Psi_P}
\ket{\Psi} = \int\int \psi_{\hbar}(\bold{p})\tilde{g}_{\beta}(\bold{p}'-\bold{p}) \ket{\bold{p} \, \bold{p}'} {\rm d}^{d}{\rm p}{\rm d}^{d}{\rm p}' \, ,
\end{eqnarray}
where 
\begin{eqnarray} \label{dB-1}
\tilde{\psi}_{\hbar}(\bold{p}) = \left(\frac{1}{\sqrt{2\pi\hbar}}\right)^d \int \psi(\bold{x}) e^{-\frac{i}{\hbar}\bold{p}.\bold{x}}{\rm d}^{d}{\rm x} \, , 
\end{eqnarray}
as in canonical QM, and
\begin{eqnarray} \label{}
\tilde{g}_{\beta}(\bold{p}'-\bold{p}) = \left(\frac{1}{\sqrt{2\pi\beta}}\right)^d \int g(\bold{x}'-\bold{x}) e^{-\frac{i}{\beta}(\bold{p}'-\bold{p}).(\bold{x}'-\bold{x})}{\rm d}^{3}{\rm x}' \, ,  
\nonumber
\end{eqnarray}
where $\beta \ll \hbar$ is a new fundamental quantum of action. 
In the smeared space model $\beta$, rather than $\hbar$, determines the quantum properties of the background geometry. 
The latter is treated as a quantum reference frame (QRF) \cite{Giacomini:2017zju} which allows existing no-go theorems for the existence of multiple Planck's constants to be circumvented \cite{Sahoo2004}. 
(The interested reader is referred to \cite{Lake:2018zeg,Lake:2019nmn,Lake2020-2} for a fuller discussion of this point.). 

Note that, in Eq. (\ref{Psi_P}), the basis $\ket{\bold{p} \, \bold{p}'}$ is entangled and cannot be separated into a simple tensor product, i.e., $\ket{\bold{p} \, \bold{p}'} \neq \ket{\bold{p}} \otimes \ket{\bold{p}'}$. 
We emphasise this by not writing a comma in between $\bold{p}$ and $\bold{p}'$, by contrast with the position space basis, $\ket{\bold{x},\bold{x}'} = \ket{\bold{x}} \otimes \ket{\bold{x}'}$. 
Nonetheless, $\tilde{g}_{\beta}(\bold{p}'-\bold{p})$ can be interpreted as the probability amplitude for the transition $\bold{p} \leftrightarrow \bold{p}'$ in smeared momentum space, by analogy with the position space representation \cite{Lake:2018zeg,Lake:2019nmn,Lake2020-2}. 

The consistency of Eqs. (\ref{|Psi>_position_space}) and (\ref{Psi_P}) requires
\begin{eqnarray} \label{mod_dB}
\braket{\bold{x},\bold{x}'|\bold{p} \, \bold{p}'} = \left(\frac{1}{2\pi\sqrt{\hbar\beta}}\right)^d e^{\frac{i}{\hbar}\bold{p}.\bold{x}} e^{\frac{i}{\beta}(\bold{p}'-\bold{p}).(\bold{x}'-\bold{x})} \, ,
\end{eqnarray} 
which is equivalent to implementing the modified de Broglie relation
\begin{eqnarray} \label{mod_dB*}
\bold{p}' = \hbar\bold{k} + \beta(\bold{k}'-\bold{k}) \, .
\end{eqnarray}
This holds for particles propagating in the smeared background and the non-canonical term may be interpreted, heuristically, as an additional momentum `kick' induced by quantum fluctuations of the geometry \cite{Lake:2018zeg,Lake:2019nmn,Lake2020-2}. 
Next, we fix the value of $\beta$ from physical considerations and show how it is related to the minimum length and momentum scales. 

The general properties of the Fourier transform \cite{Pinsky} ensure that the `wave-point' uncertainty relation, 
\begin{eqnarray} \label{Beta_UP}
\Delta_{g} x'^{i} \Delta_{g} p'_{j} \geq \frac{\beta}{2} \delta^{i}{}_{j} \, , 
\end{eqnarray}
holds in addition to Eq. (\ref{X_uncertainty}) and the HUP (\ref{HUP}), and that the inequality is saturated for Gaussian distributions. 
We may then identify the standard deviation of $|\tilde{g}_{\beta}(\bold{p})|^2$ with the $D$-dimensional de Sitter momentum,
\begin{equation} \label{sigma_g*}
\Delta_{g}p'_{j} = \tilde{\sigma}_{gj} = \frac{1}{2}m^{(D)}_{\rm dS}c \, ,
\end{equation} 
which yields the definition of $\beta$:
\begin{eqnarray} \label{beta}
\beta := (2/d)\sigma_{g}^{i}\tilde\sigma_{gi} = 2^{\frac{1}{D-2}}d^{-1} l^{(D)}_{\rm Pl} m^{(D)}_{\rm dS} \, . 
\end{eqnarray} 
In $(3+1)$ spacetime dimensions, Eq. (\ref{beta}) gives
\begin{eqnarray} \label{beta_mag}
\beta = 2\hbar\sqrt{\frac{\rho_{\Lambda}}{\rho_{\rm Pl}}} \simeq \hbar \times 10^{-61} \, , 
\end{eqnarray}
where $\rho_{\rm Pl} \simeq 10^{93}  \, {\rm g \, . \, cm^{-3}}$ is the Planck density and $\rho_{\Lambda} = \Lambda c^2/(8\pi G) \simeq 10^{-30} \, {\rm g \, . \, cm^{-3}}$ is the observed dark energy density \cite{GR_book}.

By analogous reasoning to that presented above, the probability of obtaining the observed value `$\bold{p}'$' from a smeared momentum measurement is
\begin{eqnarray} \label{EQ_PPRIMEDENSITY}
\frac{{\rm d}^{d}P(\bold{p}{\, '} | \tilde{\Psi})}{{\rm d}{\rm p}'^{d}} = \int |\tilde{\Psi}(\bold{p},\bold{p}')|^2 {\rm d}^{d}{\rm p} = (|\tilde{\psi}_{\hbar}|^2 * |\tilde{g}_{\beta}|^2)(\bold{p}') \, ,
\end{eqnarray}
which gives rise to the momentum space GUR
\begin{eqnarray} \label{P_uncertainty}
(\Delta_\Psi P_{j})^2 = (\Delta_\psi p'_{j})^2 + (\Delta_g p'_j)^2 \, .
\end{eqnarray} 
This can be obtained from the standard braket construction $(\Delta_\Psi P_{j})^2 = \braket{\Psi |(\hat{P}_{i})^{2}|\Psi} - \braket{\Psi|\hat{P}_{j}|\Psi}^2$ using
\begin{eqnarray} \label{P_operator}
\hat{P}_{j} = \int p'_{j} \, {\rm d}^{d}\hat{\mathcal{P}}_{\bold{p}{\, '}} \, ,
\end{eqnarray}
where ${\rm d}^{d}\hat{\mathcal{P}}_{\bold{p}'} = \left(\int \ket{\bold{p} \, \bold{p}'}\bra{\bold{p} \, \bold{p}'} {\rm d}^{d}{\rm p}\right){\rm d}^{d}{\rm p}'$ is the generalised projector in momentum space \cite{Lake:2018zeg}. 
Substituting the HUP (\ref{HUP}) into Eq. (\ref{P_uncertainty}) and Taylor expanding to first order then gives
\begin{eqnarray} \label{smeared_EUP}
\Delta_\Psi P_{j} \gtrsim \frac{\hbar}{2\Delta_{\psi} x'^{i}}\delta^{i}{}_{j}\left[1 + \eta(\Delta_{\psi} x'^{i})^2\right] \, ,
\end{eqnarray}
where $\eta = (1/2)(l_{\rm dS}^{(D)})^{-2}$  \cite{Lake:2018zeg}. 
For $\Delta_{\psi} p'_{j} \gg \Delta_{g} p'_{j} \simeq m_{\rm dS}c$, we have $\Delta_\Psi P_{j} \simeq \Delta_{\psi} p'_{j}$ (\ref{P_uncertainty}) so that, in this limit, Eq. (\ref{smeared_EUP}) reduces to the EUP (\ref{EUP}) when $D = 3+1$. 

Having obtained both the GUP and EUP from the smeared space formalism, we now show how they can be combined to derive the EGUP. 
Combining Eqs. (\ref{X_uncertainty}), (\ref{HUP}) and (\ref{P_uncertainty}), directly, gives 
\begin{eqnarray} \label{smeared-spaceEGUP-1}
(\Delta_{\Psi} X^{i})^2 (\Delta_{\Psi} P_{j})^2 &\geq& (\hbar/2)^2(\delta^{i}{}_{j})^2 + (\Delta_{g}x'^{i})^2(\Delta_{\Psi} P_{j})^2 
\nonumber\\
&+& (\Delta_{\Psi} X^{i})^2(\Delta_{g} p'_{j})^2 
\nonumber\\
&-& (\Delta_{g}x'^{i})^2(\Delta_{g} p'_{j})^2 \, .
\end{eqnarray}
Substituting for $\Delta_{g}x'^{i}$ and $\Delta_{g}p'_{j}$ from Eqs. (\ref{sigma_g}) and (\ref{sigma_g*}), taking the square root and expanding to first order, then ignoring the subdominant term of order $\sim l_{\rm Pl}^{(D)}m_{\rm dS}^{(D)}c$, yields
\begin{eqnarray} \label{smeared-spaceEGUP-2}
\Delta_{\Psi} X^{i} \Delta_{\Psi} P_{j} \gtrsim \frac{\hbar}{2}\delta^{i}{}_{j}\left[1 + \alpha(\Delta_{\Psi} P_{j})^2 + \eta(\Delta_{\Psi} X^{i})^2\right] \, .
\end{eqnarray}
This is equivalent to the heuristic EGUP (\ref{EGUP}) but with $\Delta x^i$ and $\Delta p_j$ replaced by well defined standard deviations, $\Delta_{\Psi} X^{i}$ and $\Delta_{\Psi} P_{j}$. 
These represent the width of the composite matter-plus-geometry state $\ket{\Psi}$ in the position and momentum space representations, respectively \cite{Lake:2018zeg,Lake:2019nmn,Lake2020-2}. 
Hence, the smeared-space formulation of the EGUP allows us to analyse the path of a QM particle in both spaces.   

Furthermore, it is straightforward to show that the product of generalised uncertainties, $\Delta_\Psi X^{i} \Delta_\Psi P_{j}$, is minimised when $\Delta_\psi x'^{i}$ and $\Delta_\psi p'_{j}$ take the values
\begin{equation} \label{EQ_CAN_DX_OPT}
(\Delta_\psi x'^{i})_{\rm opt} = \sqrt{\frac{\hbar}{2} \frac{\Delta_{g} x'^{i}}{\Delta_{g} p'_{i}}} \, , \quad (\Delta_\psi p'_{j})_{\rm opt} = \sqrt{\frac{\hbar}{2} \frac{\Delta_{g} p'_{j}}{\Delta_{g} x'^{j}}} \, , 
\end{equation}
yielding
\begin{eqnarray} \label{DXDP_opt} 
\Delta_\Psi X^{i} \, \Delta_\Psi P_{j} & \ge & \frac{(\hbar + \beta)}{2} \, \delta^{i}{}_{j} \, .
\end{eqnarray}
This result can also be obtained directly from the Schr{\" o}dinger--Robertson relation, $\Delta_{\Psi}O_1\Delta_{\Psi}O_2 \geq (1/2)\braket{\Psi | [\hat{O}_1,\hat{O}_2] |\Psi}$, since the commutator of the generalised position and momentum observables is simply a rescaled version of the canonical position-momentum commutator, with $\hbar \rightarrow \hbar + \beta$:
\begin{equation} \label{[X,P]}
[\hat{X}^{i},\hat{P}_{j}] = i(\hbar + \beta)\delta^{i}{}_{j} \, \hat{\openone} \, .
\end{equation}
The remaining commutators of the model are
\begin{equation} \label{XX_PP_commutators}
[\hat{X}^{i},\hat{X}^{j}] = 0 \, , \quad [\hat{P}_{i},\hat{P}_{j}] = 0 \, .
\end{equation}

Equations (\ref{[X,P]}) and (\ref{XX_PP_commutators}) show that GURs, including the GUP, EUP and EGUP, may be obtained without non-canonical modifications of the Heisenberg algebra \cite{Lake:2018zeg,Lake:2019oaz,Lake:2019nmn,Lake2020-1,Lake2020-2}. (See also \cite{Singleton2020,Bishop:2022des} for a similar result.) 
This allows the smeared space model to evade the problems that plague existing modified commutator models, including violation of the equivalence principle, the reference frame-dependence of the `minimum' length, and the soccer ball problem for multi-particle states \cite{Lake:2018zeg,Lake:2019oaz,Lake2020-2}, though it is worth noting that these advantages come with corresponding loss of phenomenological freedom
\footnote{For example, in this approach to GURs, we do not have the freedom to choose negative parameters for either the GUP or the EUP. 
This is in stark contrast to approaches based on modified commutators. 
(See \cite{Jizba:2009qf,Ong:2018zqn,Buoninfante:2019fwr,Kanazawa:2019llj,Scardigli:2019pme,Petruzziello:2020een,Jizba:2022icu,Lambiase:2022xde} and references therein for applications of GUP and EUP models with negative parameters to open problems in cosmology and astrophysics.)
The reason is that the non-canonical terms are derived from the standard deviations of probability distributions, in this case, $|g({\bf x}’-{\bf x})|^2$ for the GUP and $|\tilde{g}_{\beta}({\bf p}’ - {\bf p})|^2$ for the EUP. Therefore, they are positive by definition and construction. 
This remains true, even in anti-de Sitter space, in which we must set $\tilde{\sigma}_g \simeq \hbar/\sqrt{-\Lambda}$, as opposed to $\tilde{\sigma}_g \simeq \hbar/\sqrt{\Lambda}$ for an asymptotically de Sitter Universe. 
In short, it is not possible to derive negative GUP or EUP parameters from the spatial `smearing' described in our model. 
As an immediate corollary we see that, if negative GUP or EUP parameters become strongly favoured by observational data, even if their exact values remain only loosely bounded, then the smeared space GUR model will be effectively ruled out.}.

Finally, before concluding our review of the smeared space formalism, we note that the model has important implications for the description of measurement in quantum mechanics. 
Clearly, any implications of this kind are relevant to the experimental scheme for resolving the particle path, proposed in \cite{Abbott:1979bh}. 
We now illustrate these by considering a generalised position measurement, in detail. 

Applying the generalised position operator $\hat{\bold{X}}$ (\ref{X_operator}) to an arbitrary pre-measurement state $\ket{\Psi}$ returns a random measured value, $\bold{x}'$, and projects the state in the fixed background subspace of the tensor product onto
\begin{eqnarray} \label{X_measurement}
\ket{\psi_{\bold{x}'}} = \frac{1}{(|\psi|^2*|g|^2)(\bold{x}')}\int \psi(\bold{x}) g(\bold{x}'-\bold{x}) \ket{\bold{x}} {\rm d}^d\bold{x} \, , 
\end{eqnarray}
with probability $(|\psi|^2*|g|^2)(\bold{x}')$ \cite{Lake:2018zeg,Lake2020-2}.
The total state is then $\ket{\psi_{\bold{x}'}}\otimes \ket{\bold{x}'}$, which is non-normalisable, and therefore unphysical. 
This is analogous to the action of the canonical position measurement operator on $\ket{\psi}$ which projects onto the unphysical state $\ket{\bold{x}}$ with probability $|\psi(\bold{x})|^2$. 

However, in the smeared space formalism, we must reapply the fundamental `smearing' map (\ref{smearing_map}) to complete our description of the measurement process \cite{Lake:2018zeg,Lake:2019nmn,Lake2020-2}. 
In this way, a smeared measurement is split into two parts. 
In the first, a perfect projective measurement is performed on the second subspace of the tensor product, yielding the observed value of the position, $\bold{x}'$. 
Re-applying the map (\ref{smearing_map}) to $\ket{\psi_{\bold{x}'}}$ (\ref{X_measurement}) then re-smears the ket $\ket{\bold{x}}$, giving rise to a normalised post-measurement state with finite width $\Delta_{\Psi}X^{i} \gtrsim \sigma_g^{i}$ \cite{Lake:2018zeg,Lake:2019nmn,Lake2020-2}. 
 
Hence, although smeared space measurements yield precise measured values, the post-measurement states are always physical, with well defined norms. 
Their position uncertainties, which may be determined by performing multiple measurements on ensembles of identically prepared states, never fall below the fundamental smearing scale, $\sigma_g^{i} \simeq l_{\rm Pl}^{(D)}$. 
Analogous considerations hold for generalised momentum measurements, with the corresponding minimum uncertainty $\sigma_{gj} \simeq m_{\rm dS}^{(D)}c$. 

In this section, we have presented only a brief overview of the smeared space formalism. 
The interested reader is referred to references \cite{Lake:2018zeg,Lake:2019nmn,Lake2020-2} for further details.

\subsection{Important differences between smeared space and previous minimum-length models} \label{Sec.3.2} 

We now consider how the unique features of the smeared space model, vis-{\` a}-vis previous models of nonlocal geometry, affect the distribution of particle paths. 
Our analysis is qualitative but sets the stage for the quantitative analysis that follows in Sec. \ref{Sec.3}. 
The important points may be summarised as follows:

\begin{itemize}

\item By introducing degrees of freedom corresponding explicitly to the spatial background we introduce quantum paths for spatial `points', in addition to the usual paths of quantum particles on a fixed classical geometry.

\item The paths of spatial points are characterized by the smearing function, $g$, in the same way that $\psi$ determines the paths of point-particles on a fixed background. 
         (Note, however, that the quantisation scales for matter and geometry differ significantly.)

\item The resulting smeared geometry may be interpreted as an infinite superposition of Euclidean spaces in which each individual space differs from the original classical background by pair-wise exchanges of points, 
         $\bold{x} \leftrightarrow \bold{x}{\, '}$. 
         Coherent transitions between pairs of points then introduce additional stochastic fluctuations in the motion of material particles propagating in the smeared background. 

\item The net motion of a particle in the smeared geometry is therefore determined by both $\psi$ and $g$. It is the net result of two sets of paths. 
         The first are the paths the particle would have had in classical Euclidean space due to canonical quantum diffusion. The second represent additional stochastic fluctuations in position due to the relative motion of `points' in the spatial 
         background. 

\item The smearing scales, $\sigma_g$ and $\tilde{\sigma}_g$, determine the width of the smearing function in the position and momentum space representations, respectively. 
         These represent the characteristic diffusion scales for delocalised `points' in the phase space of the theory. 
         In order to recover the expected quantum gravity phenomenology, namely, the GUP and EUP, we set $\sigma_g \simeq l^{(D)}_{\rm Pl}$ and $\tilde{\sigma}_g \simeq m^{(D)}_{\rm dS}c$, where $l_{\rm Pl}^{(D)}$ is the Planck  
         length and $m_{\rm dS}^{(D)}$ is the de Sitter mass in $D$ spacetime dimensions
         \footnote{Note that, unless $g$ has compact support, the additional diffusion induced by smearing the geometry can occur over any length- or momentum-scale. However, the associated probability 
         amplitudes are vanishingly small for paths with lengths much larger than $l_{\rm Pl}^{(D)}$ in real space, or much larger than $m_{\rm dS}^{(D)}c$ in momentum space, so that these do not 
         contribute significantly to the overall motion.}.

\end{itemize}

The operational definition of a particle path in canonical QM, as determined by the measurement scheme devised in \cite{Abbott:1979bh}, is illustrated in Fig. 1. 
The operational definition of a particle path in the smeared background is illustrated in Fig. 2, for comparison. 
Unlike GUP models based on modified commutators, the presence of smearing affects the observed characteristics of the path in two ways. 
First, it alters the asymptotic properties of the path, which are obtained in the late-time limit, by introducing additional fluctuations in position and momentum  \cite{Lake:2018zeg,Lake:2019nmn,Lake2020-2}. 
Second, these fluctuations affect the resolution scale, `smearing' a sharp sphere of radius $\Delta x$ into a fuzzy sphere with average radius $\Delta X = \sqrt{(\Delta x)^2 + \sigma_g^2}$. 
In modified commutator models, the second effect does not occur and the resolution scale remains sharp, even if it is bounded from below by the minimum length \cite{Nicolini:2010dj,NiednerThesis}. 

\begin{figure}[!t] 
	\centering
	\includegraphics[width=8.7cm]{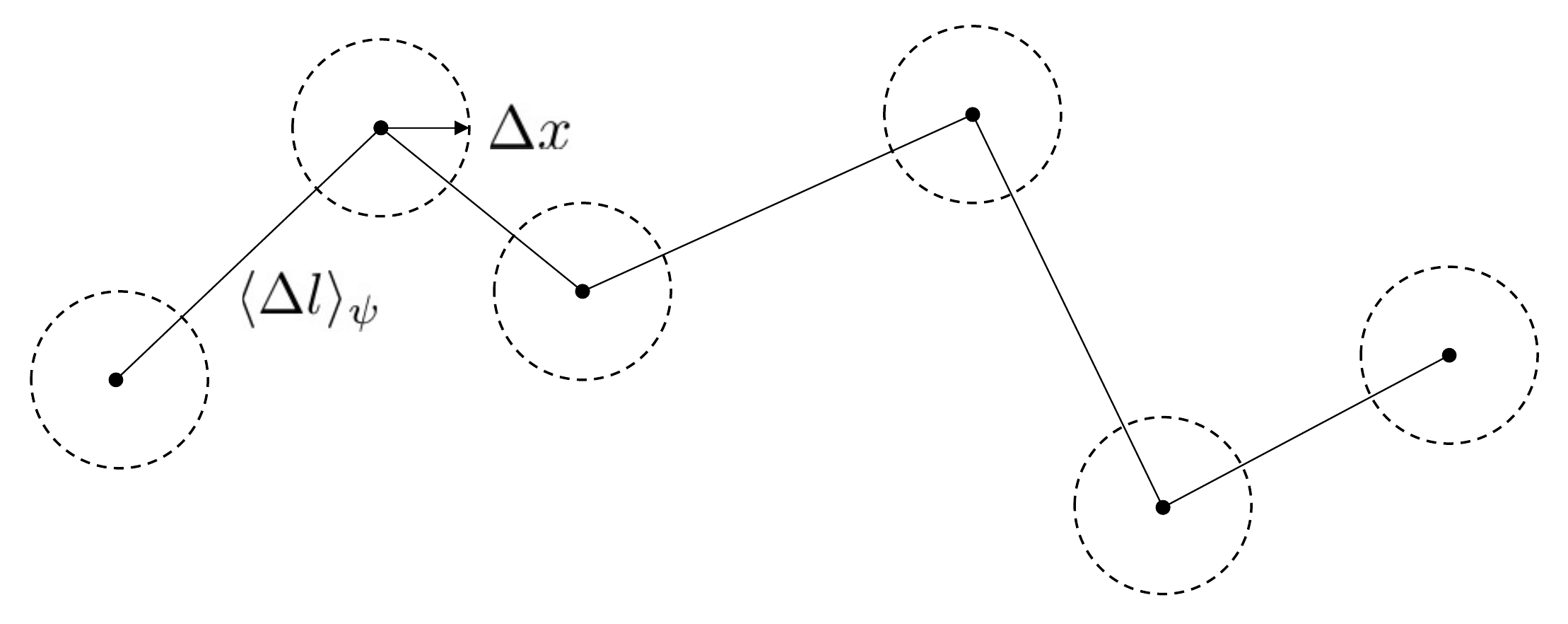}
	\caption{\label{Fig-1} Operational definition of a quantum mechanical path, in canonical QM, in the limit $\braket{\Delta l}_{\psi}  > \Delta x$. Note that, because the background space is fixed and classical, the localisation radius $\Delta x$ is 
	sharp. This remains the case in standard minimum length models, based on modified commutation relations, in which modification of the momentum space volume measure imposes the condition $\Delta x \geq l$. Taken from \cite{NiednerThesis}.}
\end{figure}

\begin{figure}[!t] 
	\centering
	\includegraphics[width=8.7cm]{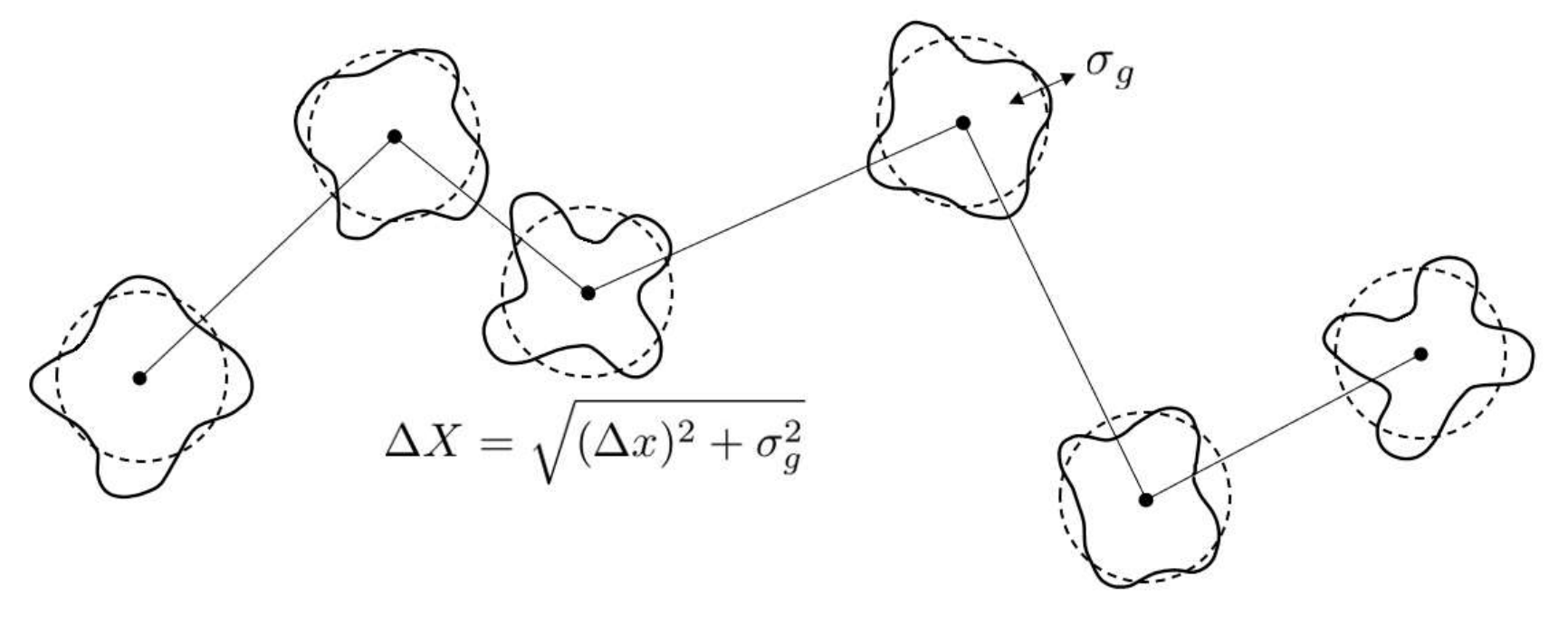}
	\caption{\label{Fig-2} Operational definition of a quantum mechanical path, in the smeared space model, in the limit $\braket{\Delta L}_{\Psi}  > \Delta X = \sqrt{(\Delta x)^2 + \sigma_g^2}$. Because the background on which the particle wave function lives is no longer classical, the localisation volume undergoes stochastic fluctuations due to the relative motion of delocalised spatial `points'. As a result, its average radius cannot drop below the smearing scale, $\sigma_g \simeq l_{\rm Pl}^{(D)}$, which corresponds to the the characteristic diffusion length for a delocalised `point' in the smeared background. This is analogous to the Compton wavelength in the matter sector, $\lambda_{\rm C}(m) = \hbar/(mc)$, which gives the characteristic diffusion scale for a point-particle propagating in a fixed classical geometry. Note that this heuristic diagram is inadequete in an important respect: the surface of the fluctuating resolution-volume is depicted as smooth, whereas, in fact, it too is fractal-like, according to our model. In a more realistic depiction, zooming in on a small section of the surface would show its discontinuous, fractal, nature. (See Fig 7.1 in Feynman and Hibbs \cite{Feynman_Hibbs}, for comparison.)}
\end{figure}

\section{The fractal properties of particle paths in smeared space} \label{Sec.4} 

In this section, we determine the fractal properties of a particle path in smeared space, which incorporates both minimum length and momentum scales. 
To do this, we must first consider the time evolution of the composite state $\ket{\Psi}$ that determines the effects of the nonlocal geometry on the canonical quantum diffusion. 

Unfortunately, the dynamics of the smeared space model are not known with certainty since it is not clear how the smearing function $g(\bold{x}'-\bold{x})$, which was assumed to be static in the generalised-measurement formalism developed in \cite{Lake:2018zeg,Lake:2019nmn,Lake2020-2}, should evolve if generalised further to depend on the `smeared time' coordinate $t'-t$. 
However, this may simply be an artifact of the nonrelativistic regime. 
We recall that in canonical QM $t$ is a parameter rather than a genuine quantum observable. 
For this reason, it cannot be smeared in the same way as $\bold{x}$, by using a map of the form (\ref{smearing_map}). 
Nonetheless, as we will now show, we may make reasonable physical assumptions that allow us to model the time evolution of the composite matter-plus-geometry state. 
Though not yet a fundamental theory, the resulting phenomenological model is compatible with canonical QM in the limit $\Delta_{\psi}x^{i}(t) \gg \sigma_g^{i}$, $\Delta_{\psi}p_{j}(t) \gg \tilde{\sigma}_{gj}$, and with the GUP, EUP and EGUP for $\Delta_{\psi}x^{i}(t) \simeq \sigma_g^{i}$, $\Delta_{\psi}p_{j}(t) \simeq \tilde{\sigma}_{gj}$. 

We begin by assuming that, in the relativistic regime, the total energy of the composite particle-plus-background system is
\begin{eqnarray} \label{}
E = \sqrt{(\bold{p}'+\bold{p}'_{\rm recoil})^2c^2 + m^2c^4} \, . 
\end{eqnarray}
Here, $m$ is the mass of the particle and $\bold{p}' = \hbar\bold{k} + \beta(\bold{k}'-\bold{k})$ (\ref{mod_dB*}) is its observed momentum, which includes the additional momentum `kicks' provided by fluctuations of the geometry. 
Conservation of momentum then implies the existence of a recoil term, $\bold{p}'_{\rm recoil} = -\beta(\bold{k}'-\bold{k})$. 
This is the additional momentum carried by the background, as a result of it imparting the `kick' $\beta(\bold{k}'-\bold{k})$ to a material body. 

The total energy of the particle-plus-geometry system is then $E =  \sqrt{\bold{p}^2c^2 + m^2c^4}$, where $\bold{p} = \hbar\bold{k}$ is the canonical particle momentum. 
In this scenario, the observed energy of a non-relativistic free quantum particle, $E'$, is given by the smeared Hamiltonian
\begin{eqnarray} \label{}
\hat{\mathcal{H}} = \frac{\hat{\bold{P}}^2}{2m} \, ,  
\end{eqnarray}
where the components of $\hat{\bold{P}}$ are given by Eq. $(\ref{P_operator})$, but the time evolution of the composite particle-plus-geometry state is generated by the canonical Hamiltonian, $\hat{H} = \hat{\bold{p}}^2/(2m)$. 
This observation may be reconciled with the formalism presented in Sec. \ref{Sec.3} by performing a unitary change of basis \cite{Lake:2019nmn,Lake2020-2}, 
\begin{eqnarray} \label{new_basis}
\braket{\bold{x},\bold{x}' | \bold{p} \, \bold{p}'} \mapsto \braket{\bold{x}' - \bold{x} | \bold{p}' - \bold{p}} \, . 
\end{eqnarray}
In this basis the smeared state (\ref{|Psi>_position_space}) / (\ref{Psi_P}) becomes separable yielding $\ket{\Psi} = \ket{\psi} \otimes \ket{g}$, where $\ket{g} = \int g(\bold{x}'-\bold{x}) \ket{\bold{x}' - \bold{x}}{\rm d}^dx'$ is the quantum state associated with the whole background geometry, rather than a single delocalised `point', i.e., $\ket{g} \neq \ket{g_{\bold{x}}}$ \cite{Lake:2019nmn,Lake2020-2}.

In the new basis (\ref{new_basis}) the Hamiltonian that drives the time-evolution of the system then takes the form $\hat{\bold{p}}^2/(2m) \otimes \hat{\openone}$. 
This implies that $\ket{\psi}$ evolves according to the laws of canonical QM whereas $\ket{g}$ does not evolve in time. 
This model treats the smearing introduced in Eq. (\ref{smearing_map}) as a simple fact of nature: in this scenario, there are no true points in the quantum geometry, only delocalised `points', whose associated quantum amplitudes effectively smear them over Planck-sized volumes. 

The form of $\ket{\Psi_t}$ is therefore determined as follows. 
First, we take an initial canonical QM state $\ket{\psi_{0}} = \int \psi_{0}(\bold{x})\ket{\bold{x}}{\rm d}^{d}x$ and evolve this using the canonical time-evolution operator (\ref{U(t)}), giving $\ket{\psi_{t}} = \hat{U}(t)\ket{\psi_{0}}$. 
We then set $\ket{\Psi_t} = \ket{\psi_{t}} \otimes \ket{g}$, with $\ket{g} = \int g(\bold{x}'-\bold{x}) \ket{\bold{x}' - \bold{x}}{\rm d}^dx'$, before performing the inverse basis change $\braket{\bold{x}' - \bold{x} | \bold{p}' - \bold{p}} \mapsto \braket{\bold{x},\bold{x}' | \bold{p} \, \bold{p}'}$ \cite{Lake:2019nmn,Lake2020-2}. 
This gives 
\begin{eqnarray} \label{}
\ket{\Psi_t} &=& \int \psi_{t}(\bold{x})g(\bold{x}'-\bold{x})\ket{\bold{x},\bold{x}'}{\rm d}^dx{\rm d}^dx' 
\nonumber\\
&=& \int \tilde{\psi}_{t}(\bold{p})\tilde{g}_{\beta}(\bold{p}'-\bold{p})\ket{\bold{p} \, \bold{p}'}{\rm d}^dp{\rm d}^dp' \, ,
\end{eqnarray}
where $\psi_{t}(\bold{x})$ is obtained by solving the canonical time-dependent Schr{\" o}dinger equation. 
Here, we have omitted the subscript $\hbar$ on $\tilde{\psi}$, for the sake of notational simplicity.

Physically, this is equivalent to assuming that the canonical quantum state $\ket{\psi}$, which is defined only with respect to a fixed classical background, evolves according to the usual laws of canonical QM in each Euclidean space embedded within the smeared superposition of geometries. 
The additional fluctuations in position and momentum, which give rise to the GUP, EUP and EGUP, are due entirely to coherent transitions of the form $\bold{x} \leftrightarrow \bold{x}'$, $\bold{p} \leftrightarrow \bold{p}'$, induced by smearing. 
The probability densities for these transitions, $|g(\bold{x}'-\bold{x})|^2$ and $|\tilde{g}_{\beta}(\bold{p}'-\bold{p})|^2$, are independent of $|\psi(\bold{x})|^2$ and $|\tilde{\psi}(\bold{p})|^2$. 
In this sense, the back-reaction of the quantum matter on the quantum geometry is neglected. 

\subsection{The position space representation} \label{Sec.4.1} 

We now have all the tools we need to determine the fractal properties of the path of a free particle in the smeared space model. 
The path length traversed in time $\Delta t$ is defined as
\begin{eqnarray} \label{<Delta_L>_{Psi}-1}
\braket{\Delta L}_{\Psi} = \braket{\Psi | \hat{U}^{\dagger}(\Delta t) \hat{\bold{X}}^2 \hat{U}(\Delta t) |\Psi}^{1/2} \, ,
\end{eqnarray}
where $\hat{U}(\Delta t)$ is given by Eq. (\ref{U(t)}), $\ket{\Psi}$ is given by Eqs. (\ref{|Psi>_position_space}) and (\ref{Psi_P}), and $\hat{\bold{X}}$ is given by Eq. (\ref{X_operator}). 
The total path length traversed in $N$ steps, corresponding to the total time interval $T = N\Delta t$, is 
\begin{eqnarray} \label{<L>_{Psi}-1} 
\braket{L}_{\Psi} = N \braket{\Delta L}_{\Psi} \, . 
\end{eqnarray}

Taking $|\psi_0(\bold{x})|^2$ to be a Gaussian with width $\sigma_{\psi}(0) = \Delta x$, as in Eq. (\ref{psi_0}), and setting 
\begin{eqnarray} \label{|g(x'-x)|^2}
|g(\bold{x}'-\bold{x})|^2 = \left(\frac{1}{\sqrt{2\pi} \sigma_g}\right)^{d} \exp\left[-\frac{(\bold{x}'-\bold{x})^2}{2\sigma_g^2}\right] \, ,
\end{eqnarray}
then gives  
\begin{eqnarray} \label{<Delta L>_{Psi}-2}
&&\braket{\Delta L}_{\Psi} = \sqrt{\sigma^2_{\psi}(\Delta t) + \sigma_g^2}
\nonumber\\
&=& \frac{\hbar}{2m}\frac{\Delta t}{\Delta x} \sqrt{1 + \frac{4m^2}{\hbar^2}\frac{(\Delta x)^2[(\Delta x)^2 + \sigma_g^2]}{(\Delta t)^2}} \, . 
\end{eqnarray} 
Here, $(\Delta x)^d$ is the position uncertainty that the experimenter aims for but is unable to achieve with absolute precision. 
Instead, the particle is localised to within a region of space with average volume $(\Delta X)^d$, where
\begin{eqnarray} \label{Delta_X}
\Delta X = \sqrt{(\Delta x)^2 + \sigma_g^2} \, .
\end{eqnarray}

The experimentalist has no control over $\sigma_g$, which we assume is given by Eq. (\ref{sigma_g}), but their choice of $\Delta x$ effectively fixes the average uncertainty in each meaurement to $\Delta X = \Delta_{\Psi}X(0)$. 
We therefore define the smeared-space Hausdorff length as 
\begin{eqnarray} \label{<L_H>_{Psi}-1}
\braket{L_{\rm H}}_{\Psi} = \braket{L}_{\Psi} \, . \, (\Delta X)^{D_{\rm H}-1} \, . 
\end{eqnarray}
This can be written in terms of $\Delta x$, $\Delta t$ and $\sigma_g$ as
\begin{eqnarray} \label{<L_H>_{Psi}-2}
&&\braket{L_{\rm H}}_{\Psi} = \frac{N\hbar}{2m}\frac{\Delta t}{\Delta x} \times
\nonumber\\
&& \sqrt{1 + \frac{4m^2}{\hbar^2}\frac{(\Delta x)^2[(\Delta x)^2 + \sigma_g^2]}{(\Delta t)^2}} [(\Delta x)^2+\sigma_g^2]^{D_{\rm H}-1} 
\end{eqnarray}
or, equivalently, in terms of $\Delta X$, $\Delta t$ and $\sigma_g$ as
\begin{eqnarray} \label{<L_H>_{Psi}-3}
&&\braket{L_{\rm H}}_{\Psi} = \frac{N\hbar}{2m}\frac{\Delta t}{\sqrt{(\Delta X)^2-\sigma_g^2}} \times
\nonumber\\
&& \sqrt{1 + \frac{4m^2}{\hbar^2}\frac{(\Delta X)^2[(\Delta X)^2 - \sigma_g^2]}{(\Delta t)^2}} \, . \, (\Delta X)^{D_{\rm H}-1} . 
\end{eqnarray}

Treating $\Delta x$, $\Delta t$ and $\sigma_g$ as independent variables in Eq. (\ref{<L_H>_{Psi}-2}) and imposing the condition $d\braket{L_{\rm H}}_{\Psi}/d(\Delta x) = 0$ yields the polynomial equation
\begin{eqnarray} \label{polynomial-1_smeared}
&&(\Delta x)^{D_{\rm H}-2} + \frac{(D_{\rm H}-2)}{D_{\rm H}} \frac{\hbar^2}{4m} (\Delta t)^2 (\Delta x)^{D_{\rm H}-6} 
\nonumber\\
&+& \sigma_g^2\left[(\Delta x)^{D_{\rm H}-2} - \frac{\hbar^2}{4m} \frac{(\Delta t)^2}{D_{\rm H}}(\Delta x)^{D_{\rm H}-8}\right] = 0 \, .
\end{eqnarray}
Immediately we see that, in the limit $\Delta t \rightarrow \infty$, this reduces to the condition $D_{\rm H} \simeq 2 + (\sigma_g/\Delta x)^2$, unlike Eq. (\ref{polynomial-1}). 
However, this is also a contradiction, since (\ref{polynomial-1_smeared}) was derived under the assumption that $D_{\rm H}$ is independent of $\Delta x$. 
As before, these results can be reconciled by setting $\Delta t = \xi \, . \, 4m(\Delta x)^2/\hbar$ (\ref{xi}) and 
\begin{eqnarray} \label{epsilon_smeared}
\sigma_g = \epsilon \, . \, \Delta x \, , 
\end{eqnarray}
by analogy with Eq. (\ref{epsilon}). 
This is equivalent to imposing $d\braket{L_{\rm H}}_{\Psi}/d(\Delta x)|_{\Delta x = \sqrt{\hbar\Delta t/4m\xi} = \sigma_g/\epsilon} = 0$, giving
\begin{eqnarray} \label{D_H_smeared}
D_{\rm H}(\xi,\epsilon) = \frac{2 + \epsilon^2}{1 + \frac{1+\epsilon^2}{4\xi^2}} \, . 
\end{eqnarray}
It is straightforward to show that, using Eq. (\ref{<L_H>_{Psi}-3}) instead of (\ref{<L_H>_{Psi}-2}), and imposing the conditions $d\braket{L_{\rm H}}_{\Psi}/d(\Delta X) = 0$ and $d\braket{L_{\rm H}}_{\Psi}/d(\Delta X)|_{\Delta X = \sqrt{\hbar^2(\Delta t)^2/16m^2\xi^2 + \sigma_g^2} = \sqrt{1+\epsilon^2}\sigma_g/\epsilon} = 0$, respectively, yields exactly the same results. 

From (\ref{D_H_smeared}) we see that, as $\xi \rightarrow \infty$, $D_{\rm H} > 2$. 
For $\xi \in [1,\infty)$, the Hausdorff dimension varies in the range $D_{\rm H} \in [(8/5)(1+\epsilon^2/2),2+\epsilon^2)$, where the lower limit is given to first order in $\epsilon^2$. 
Hence, in the smeared space model, the presence of the minimum length $\sigma_g$ always increases $D_{\rm H}$, relative to its canonical QM value. 

Moreover, Eq. (\ref{D_H_smeared}) implies
\begin{eqnarray} \label{D_H><2}
D_{\rm H} \gtrless 2 \iff \xi \gtrless \sqrt{\frac{1+\epsilon^2}{2\epsilon^2}} \, . 
\end{eqnarray} 
This is condition always satisfied when $\epsilon = 1$ since $\xi \geq 1$ (\ref{Delta t}). 
However, for $\epsilon \ll 1$, which is the most physically reasonable scenario when $\sigma_g \simeq l_{\rm Pl}^{(D)}$, $\xi$ must take increasingly large values in order to ensure that $D_{\rm H} > 2$. 
In other words, the larger the spatial resolution $\Delta x$, as compared to the minimum length $\sigma_g \simeq l_{\rm Pl}^{(D)}$, the longer it takes for the Hausdorff length of the smeared particle path to rise above the upper bound set by canonical QM. 
For $\sigma_g/\Delta x \rightarrow 0$, this bound cannot be breached, even as $\Delta t \rightarrow \infty$. 
Conversely, for $\sigma_g/\Delta x \rightarrow 1$, the canonical QM bound $D_{\rm H} = 2$ is automatically obtained by taking the minimum possible time-interval, $\Delta t = 4m(\Delta x)^2/\hbar$, while taking longer time intervals between measurements yields $D_{\rm H} > 2$. 

For clarity, the limiting values of the Hausdorff dimension $D_{\rm H}(\xi,\epsilon)$ (\ref{D_H_smeared}) are shown in Table 1 below. 
The limiting values of $\epsilon$  are given in the top row whereas the limiting values of $\xi$ are given in the first column. 

\begin{center} 
\begin{tabular}{ |c|c|c| } 
 \hline
 $\,\, \epsilon \, / \, \xi \,\,$ & 0 & \,\,\, 1 \,\,\, \\ 
  \hline
 1 & \, 1.6 \, & \,\,\, 2 \,\,\, \\ 
  \hline
 $\infty$ & \, 2 \, & \,\,\, 3 \,\,\, \\ 
 \hline
\end{tabular}
\end{center}

We note that the extremal case $D_{\rm H} = 3$ is obtained in the asymptotic limit, $\Delta t \rightarrow \infty$, when $\Delta x = \sigma_g$. 
This makes intuitive sense. 
On this scale, the self-similar structure induced by the motion of delocalised `points' in the smeared background is exactly comparable to the self-similar structure induced by the motion of the point-particle particle due to canonical quantum diffusion. 
On a fixed background, the latter increases the dimension of the path from its classical value, 1, the topological dimension of a line, to 2. 
Superimposing the addition self-similar structure induced by the diffusion of nonlocal points increases this value, in like manner, from 2 to 3. 

\subsection{The momentum space representation} \label{Sec.4.2} 

In the momentum space picture we define the path length traversed in the interval $\Delta t$ as
\begin{eqnarray} \label{<Delta_tilde_L>}
\braket{\Delta \tilde{L}}_{\Psi} &=& \braket{\Psi | \hat{U}^{\dagger}(\Delta t) \hat{\bold{P}}^2 \hat{U}(\Delta t) |\Psi}^{1/2} \, , 
\end{eqnarray} 
yielding
\begin{eqnarray} \label{<tilde_L>}
\braket{\Delta \tilde{L}}_{\Psi} = \sqrt{\tilde{\sigma}_{\psi}^2(\Delta t) + \tilde{\sigma}_{g}^2}
= \sqrt{\frac{\hbar^2}{4\sigma_{\psi}^2(\Delta t)} + \frac{\beta^2}{4\sigma_{g}^2}} \, .
\end{eqnarray} 
The total path length traversed in $T = N\Delta t$ is 
\begin{eqnarray}  \label{<tilde_L>}
\braket{\tilde{L}}_{\Psi} = N \braket{\Delta \tilde{L}}_{\Psi} \, , 
\end{eqnarray} 
and the momentum-space Hausdorff length is defined as
\begin{eqnarray}  \label{<tilde_L_H>*}
\braket{\tilde{L}_{\rm H}}_{\Psi} = \braket{\tilde{L}}_{\Psi} \, . \, (\Delta P)^{\tilde{D}_{\rm H}-1} \, , 
\end{eqnarray} 
where 
\begin{eqnarray} \label{Delta_P}
\Delta P = \sqrt{(\Delta p)^2 + \tilde{\sigma}_g^2} = \sqrt{(\Delta p)^2 + \frac{\beta^2}{4\sigma_{g}^2}} \, . 
\end{eqnarray} 

Imposing the condition $d\braket{\tilde{L}_{\rm H}}_{\Psi}/d(\Delta p) = 0$ yields a polynomial equation that can be written in the form
\begin{eqnarray} \label{polynomial-2_smeared}
&&\frac{d\sigma_{\psi}(\Delta t) }{d(\Delta p)} \sigma_{\psi}(\Delta t) (\Delta p)^{\tilde{D}_{\rm H}-1} 
\nonumber\\
&-& (\tilde{D}_{\rm H}-1)\sigma_{\psi}^2(\Delta t) (\Delta p)^{\tilde{D}_{\rm H}-2}
\nonumber\\
&+& \frac{\delta^2}{\sigma_g^2}\left\{\left(\frac{\hbar}{2}\right)^2\frac{d\sigma_{\psi}(\Delta t)}{d(\Delta p)}\frac{\sigma_{\psi}(\Delta t)}{\Delta p} - (\tilde{D}_{\rm H}-1)\sigma_{\psi}^4(\Delta t)\right\} 
\nonumber\\
&\times& (\Delta p)^{\tilde{D}_{\rm H}-2} \, , 
\end{eqnarray}
where $\sigma_{\psi}(\Delta t)$ is expressed in terms of $\Delta p$, 
\begin{eqnarray} \label{}
\sigma_{\psi}(\Delta t) = \frac{\hbar}{2\Delta p}\sqrt{1 + \frac{4(\Delta p)^4(\Delta t)^2}{m^2}} \, , 
\end{eqnarray}
and the parameter $\delta$ is defined as
\begin{eqnarray} 
\delta := \beta/\hbar \, . 
\end{eqnarray}

Using the minimum length and momentum scales given in Eqs. (\ref{sigma_g}) and (\ref{sigma_g*}) gives $\delta^2 = 4\rho_{\Lambda}/\rho_{\rm Pl} \simeq 10^{-122}$ but taking the limit $\delta \rightarrow 0$ ($\beta \rightarrow 0$), which is equivalent to setting $\sigma_g \rightarrow 0$ and $\tilde{\sigma}_g \rightarrow 0$ simultaneously, yields the canonical QM limit of the smeared space model \cite{Lake:2018zeg}. 
In this case, Eq. (\ref{polynomial-2_smeared}) reduces to Eq. (\ref{polynomial-2}), as required. 

We now parameterise the time interval between measurements such that
\begin{eqnarray}  \label{polynomial-3_smeared}
\Delta t = \xi \, . \, \frac{4m(\Delta x)^2}{\hbar} = \xi \, . \, \frac{m\hbar}{(\Delta p)^2} \, , 
\end{eqnarray}
with $\xi \gtrsim 1$, as before, so that (\ref{polynomial-2_smeared}) becomes 
\begin{eqnarray} 
&&[4\xi^2 - 1 - (\tilde{D}_{\rm H}-1)(4\xi^2 + 1)] (\Delta p)^{\tilde{D}_{\rm H}-4}
\nonumber\\
&+& \frac{\delta^2}{\sigma_g^2} \left(\frac{\hbar}{2}\right)^2 [4\xi^2 - 1 - (\tilde{D}_{\rm H}-1)(4\xi^2 + 1)^2] 
\nonumber\\
&\times& (\Delta p)^{\tilde{D}_{\rm H}-6} = 0 \, . 
\end{eqnarray}

Next, we use the fact that 
\begin{eqnarray}  \label{}
\tilde{\sigma}_g = \frac{\beta}{2\sigma_g} = \frac{\beta}{2\epsilon \Delta x} = \frac{\beta}{\hbar}\frac{\Delta p}{\epsilon} = \delta \frac{\Delta p}{\epsilon} 
\end{eqnarray}
to parameterise the resolution of the momentum measurements in terms of the minimum scale as 
\begin{eqnarray} \label{tilde_epsilon-1}
\tilde{\sigma}_g = \tilde{\epsilon} \, . \, \Delta p = (\delta/\epsilon) \, . \, \Delta p \, ,
\end{eqnarray}
where $\tilde{\epsilon} \lesssim 1$ requires $\epsilon \gtrsim \delta \simeq 10^{-61}$. 
This condition is physically reasonable since it is equivalent to setting $\Delta x \lesssim \sigma_g/\delta \simeq l^{(D)}_{\rm dS}$, where $l^{(D)}_{\rm dS}$ is the asymptotic de Sitter horizon in a $D$-dimensional universe. Substituting from Eqs. (\ref{polynomial-3_smeared}) and (\ref{tilde_epsilon-1}), Eq. (\ref{polynomial-2_smeared}) can then be solved to give
\begin{eqnarray} \label{tilde_D_H_smeared}
\tilde{D}_{\rm H}(\xi,\epsilon) = 1 + \left(\frac{4\xi^2-1}{4\xi^2+1}\right) \left[\frac{1+(\delta/\epsilon)^2}{1+(\delta/\epsilon)^2(4\xi^2+1)}\right] \, . 
\end{eqnarray}

Equation (\ref{tilde_D_H_smeared})  results from imposing the condition $d\braket{\tilde{L}_{\rm H}}_{\Psi}/d(\Delta p)|_{\Delta p = \sqrt{\xi m\hbar/\Delta t} = (\epsilon/\delta)\tilde{\sigma}_g} = 0$, or, equivalently, imposing $d\braket{\tilde{L}_{\rm H}}_{\Psi}/d(\Delta P)|_{\Delta P = \sqrt{\xi^2 m^2\hbar^2/(\Delta t)^2 + \tilde{\sigma}_g^2} = \sqrt{1+(\delta/\epsilon)^2}\tilde{\sigma}_g/(\delta/\epsilon)}$ $= 0$. 
It reduces to Eq. (\ref{tilde_D_H}) in the limit $\delta \rightarrow 0$, but for nonzero $\delta$ the factor in square brackets is always less than one, giving $\tilde{D}_{\rm H} < 2/(1+\frac{1}{4\xi^2})$. 

For clarity, the limiting values of the momentum space Hausdorff dimension, $\tilde{D}_{\rm H}(\xi,\epsilon)$ (\ref{tilde_D_H_smeared}), are shown in Table 2. 
The limiting values of $\tilde{\epsilon} = \delta/\epsilon$ are given in the top row and the limiting values of $\xi$ are given in the first column. 

\begin{center} 
\begin{tabular}{ |c|c|c| } 
 \hline
 $\,\, \tilde{\epsilon} \, / \, \xi \,\,$ & 0 & \,\,\, 1 \,\,\, \\ 
  \hline
 1 & \, 1.6 \, & \,\,\, 1 \,\,\, \\ 
  \hline
 $\infty$ & \, 2 \, & \,\,\, 1.2 \,\,\, \\ 
 \hline
\end{tabular}
\end{center}
It is immediately clear that $\tilde{D}_{\rm H} \gtrsim 1$ when $\Delta p \gtrsim \tilde{\sigma}_g$, irrespective of the chosen time-interval $\Delta t$. 
However, it is important to recognise that the extremal case $\tilde{D}_{\rm H} \simeq 1$, $\Delta p \simeq \tilde{\sigma}_g$ does not correspond to the classical regime. 
The particle path remains embedded as a fractal in the $d$-dimensional space, but its Hausdorff dimension is significantly reduced, compared to its canonical QM value, by the corrections induced by the EUP.  

\subsection{Fractal properties of the dark energy field and astrophysical black holes} \label{Sec.4.3} 

In this section, we consider the possible implications of our previous analysis for the fractal properties of the dark energy field. 

We begin by noting that, for $\xi \lesssim \epsilon/(2\delta)$, we may expand the terms in square brackets in Eq. (\ref{tilde_D_H_smeared}) as
\begin{eqnarray} \label{tilde_D_H_smeared_approx}
\tilde{D}_{\rm H}(\xi,\epsilon) \simeq \frac{2 - (\delta/\epsilon)^2\xi^2}{1+\frac{1}{4\xi^2}} \, .
\end{eqnarray} 
This condition is always valid, in any realistic experimental scenario, since it is equivalent to setting $\Delta t \lesssim 4m\Delta x \, l_{\rm dS}^{(D)}/\hbar$. 
Combining this inequality with Eq. (\ref{polynomial-3_smeared}) gives $\Delta x \lesssim \sigma_g/\delta \simeq l^{(D)}_{\rm dS}$, which ensures that $\epsilon \gtrsim \delta$ ($\Delta p \gtrsim \tilde{\sigma}_g \simeq m_{\rm dS}^{(D)}c$), as discussed above. 

Interestingly, Eqs. (\ref{D_H_smeared}) and (\ref{tilde_D_H_smeared_approx}) indicate the existence of a critical case, which occurs when $\xi \simeq \epsilon^2/(2\delta)$. 
This is equivalent to setting $\Delta t \simeq (m/\hbar)l_{\rm Pl}^{(D)}l_{\rm dS}^{(D)}$. 
For time-intervals of this duration, we have  
\begin{eqnarray} \label{critical-1}
D_{\rm H} \simeq 2 + \epsilon^2 \, \quad \tilde{D}_{\rm H} \simeq 2 - \epsilon^2 \, ,  
\end{eqnarray} 
so that the momentum space Hausdorff dimension is decreased by the same amount that the position-space Hausdorff dimension is increased. 
By contrast, both shorter and longer time-intervals introduce further asymmetry. 
For $\Delta t \lesssim (m/\hbar)l_{\rm Pl}^{(D)}l_{\rm dS}^{(D)}$, $D_{\rm H}$ is increased by the GUP more than $\tilde{D}_{\rm H}$ is decreased by the EUP, whereas the reverse is true for $\Delta t \gtrsim (m/\hbar)l_{\rm Pl}^{(D)}l_{\rm dS}^{(D)}$. 

What is the physical significance of this critical time-interval? 
To answer this question, we must return to the conditions (\ref{EQ_CAN_DX_OPT}). 
We recall that these give the position and momentum space resolutions required to minimise the product of the generalised uncertainties, $\Delta_{\Psi}X^{i}\Delta_{\Psi}P_{j}$. 
This saturates the inequality in the smeared-space GUR, giving rise to the Schr{\" o}dinger-Robertson bound (\ref{DXDP_opt}). 
It is straightforward to show that imposing Eqs. (\ref{EQ_CAN_DX_OPT}) corresponds to setting $\epsilon^2 \simeq \delta$. 
Combining this with $\xi \simeq \epsilon^2/(2\delta)$, the critical condition for symmetry between the GUP- and EUP-induced corrections to $D_{\rm H}$ and $\tilde{D}_{\rm H}$, then gives
\begin{eqnarray} \label{critical-2}
\epsilon^2 \simeq \delta \, , \quad \xi \simeq 1 \, ,
\end{eqnarray}
to within numerical factors of order unity. 

For simplicity, we restrict our attention to $(3+1)$ dimensions from here on. 
In this case, Eq. (\ref{critical-2}) implies
\begin{eqnarray} \label{critical-3}
(\Delta x)_{\rm opt} \simeq l_{\Lambda} \, , \quad (\Delta p)_{\rm opt} \simeq \frac{1}{2} m_{\Lambda}c \, , 
\end{eqnarray}
where
\begin{eqnarray} \label{critical-4}
l_{\Lambda} &=& 2^{\frac{1}{4}}\sqrt{l_{\rm Pl}l_{\rm dS}} \simeq 0.1 \, {\rm mm} \, , 
\nonumber\\
m_{\Lambda} &=& 2^{-\frac{1}{4}}\sqrt{m_{\rm Pl}m_{\rm dS}} \simeq 10^{-3} \, {\rm eV/c^2} \, ,
\end{eqnarray}
by Eqs. (\ref{sigma_g}) and (\ref{sigma_g*}). 
For $m \simeq m_{\Lambda}$, we then have 
\begin{eqnarray} \label{}
\Delta t \simeq t_{\Lambda} = l_{\Lambda}/c \, , 
\end{eqnarray}
and the associated energy density of the generalised particle wave function $\Psi$ is comparable to the dark energy density,
\begin{eqnarray} \label{critical-5}
\rho_{\Psi} = \frac{3}{4\pi}\frac{(\Delta p)_{\rm opt}}{(\Delta x)_{\rm opt}^3c} \simeq \frac{\Lambda c^2}{8\pi G} \simeq 10^{-30} \, {\rm g \, . \, cm^{-3}} \, . 
\end{eqnarray}

In \cite{Burikham:2015nma,Lake:2017ync,Lake:2017uzd,Hashiba:2018hth,Shubham2020} it was conjectured that a space-filling `sea' of weakly interacting fermions of mass $m_{\rm DE} \simeq m_{\Lambda}$, existing in a critical Hagedorn state, could give rise to the present-day accelerated expansion of the universe. 
In this model, the dark energy density is approximately constant over large distances but appears granular over length scales of order $0.1$ mm, as tentatively suggested by recent observations \cite{Perivolaropoulos:2016ucs,Antoniou:2017mhs,0.1mm_latest}. 
Here, we have shown that such a sea of dark energy particles exhibits interesting fractal properties, and corresponds to a critical case in which the GUP- and EUP-induced corrections to to the Hausdorff dimensions in the position and momentum space representations are symmetric, i.e., equal in magnitude but opposite in sign, according to Eq. (\ref{critical-1}), with $\epsilon^2 \simeq \delta \simeq 10^{-61}$. 

Practically, this is likely to be indistinguishable from the canonical quantum regime. 
Nonetheless, the existence of such a critical fractal state, corresponding to the critical energy density of the present-day universe \cite{GR_book}, hints at a deeper connection between the microscopic dynamics induced by quantum gravity corrections to canonical QM and the large-scale structure of the universe.  

Finally, before concluding this section, we briefly consider the implications of our model for another important type of astrophysical system: black holes. 
Treating quantum fluctuations of space-time `points' as random walks that gradually build up fractal structure {\it should} imply the fractalisation of the black hole horizon. 
This is akin to Barrow's notion of a fractal-like `wrinkled' black hole, leading to a modified entropy-area law, $S \propto A^{1 + \Delta/2}$, $\Delta \in [0,1]$ \cite{Barrow:2020tzx}. 
The cosmological implications of such a model, as applied to the apparent cosmological horizon, are also profound 
(see \cite{Barrow:2020tzx,Leon:2021wyx,DiGennaro:2022ykp,Farsi:2022dvd,Luciano:2022pzg,Nojiri:2021jxf} for recent works) and could help to relieve the Hubble tension \cite{Verde:2019ivm,DiValentino:2021izs}. 
 
How to derive the Barrow entropy index $\Delta$ remains an open problem in quantum gravity research and it may be hoped that, with future study, one possible form of $\delta$ could be predicted from within the smeared space model. 
In his seminal paper \cite{Barrow:2020tzx}, Barrow worked within canonical QM, though he also considered what implications GUP-style modifications of the uncertainty principle could have for his proposal, stating that ``An interesting extension of these calculations is to replace the HUP by its extension when gravitational forces are included''. 
What if, instead of {\it adding} GUP effects to a fractal Barrow entropy model, we could instead derive both the GUP and the Barrow entropy index from a single underlying formalism? 
We will address this issue in depth in a future work.

\section{Discussion} \label{Sec.5}

We have investigated the fractal properties of the path of a quantum particle in three models: canonical QM, the minimum-length model considered previously by Nicolini and Niedner \cite{Nicolini:2010dj,NiednerThesis}, and the model of quantum geometry known as `smeared space', which was recently proposed in \cite{Lake:2018zeg,Lake:2019oaz,Lake:2019nmn,Lake:2021beh,Lake2020-1,Lake2020-2}. 
In canonical QM, the fractal properties of the particle path are determined by the HUP, whereas the other two models employ GURs. 

The Nicolini-Niedner model implements the GUP by introducing modified commutation relations and a modified phase space volume. 
In this respect, it is representative of a large class of similar GUP models proposed in the existing literature \cite{Tawfik:2014zca,Tawfik:2015rva}. 
This approach allows us to analyse the effects of introducing a minimum length, but not a minimum momentum, since the momentum space representation is not well defined \cite{Nicolini:2010dj,NiednerThesis}. 
By contrast, the smeared space model allows us to analyse the effects of both minimum length and momentum scales, since both the position and momentum space pictures are well defined. 
The former gives rise to the GUP, whereas the latter gives rise to the EUP. 
These may then be combined to give the EGUP.

In this work, we have presented new results in all three formalisms. 
In canonical QM, we extended the pioneering analysis of Abbot and Wise \cite{Abbott:1979bh} in two ways. 
First, we showed that their estimate of the Hausdorff dimension of the fractal particle path, $D_{\rm H} = 2$, corresponds to the asymptotic limit $\Delta t \rightarrow \infty$. 
In this scenario, the path of the particle is sampled by series of measurements that resolve its position to within a finite volume $\sim (\Delta x)^d$, in $d$ spatial dimensions, and successive measurements are separated by very large time-intervals, $\Delta t \gg 4m(\Delta x)^2/\hbar$. 
We showed that, for smaller time-intervals, $\Delta t \gtrsim (\Delta t)_{\rm min} = 4m(\Delta x)^2/\hbar$, where the minimum bound follows from the energy-time uncertainty relation, $D_{\rm H}$ can be significantly reduced. 

This has a clear physical interpretation and is due to the fact that sampling interrupts the process of free quantum diffusion, which is akin to Brownian motion \cite{Nelson_1966}. 
The stochastic motion gradually builds up the fractal structure giving rise to self-similarity on small scales over short time-intervals and self-similarity on larger scales at later times. 
Thus, at any finite time, the quantum path-fractal is incomplete. 
This is manifested as a decrease in the Hausdorff dimension, relative to its asymptotic value of $D_{\rm H} = 2$.  

Second, we extended the analysis presented in \cite{Abbott:1979bh} by defining the Hausdorff dimension for the particle path in momentum space, $\tilde{D}_{\rm H}$. 
This makes physical sense, since any measurement that localises the particle to within a volume of order $(\Delta x)^d$ in position space also localises it to within a momentum space volume of order $(\Delta p)^d$, where $\Delta p \gtrsim \hbar/(2\Delta x)$. 
Finite-precision position measurements therefore constitute de facto momentum measurements, and vice versa. 
Our analysis showed that, in canonical QM, the paths of the particle in the position and momentum space representations are isomorphic, with equal Hausdorff dimensions, $D_{\rm H} = \tilde{D}_{\rm H}$. 

In the modified commutator model \cite{Nicolini:2010dj,NiednerThesis}, Nicolini and Niedner also analysed the late-time limit, $\Delta t \rightarrow \infty$. 
We extended their analysis by considering shorter sampling times, $\Delta t \simeq (\Delta t)_{\rm min}$, and determined the exact dependence of the Hausdorff dimension on both the sampling time-interval and the value of the minimum length $l$. 
In this way, $D_{\rm H}$ was determined to be a function of two dimensionless parameters, $\xi = \Delta t/(\Delta t)_{\rm min}$ and $\epsilon = l/\Delta x$. 
In addition, we addressed a loophole in their derivation of the relation between the Hausdorff and spectral dimensions of the GUP-modified particle path. 
Our analysis showed that their results remain valid, even though a number of subtleties must be considered, in order to derive them rigorously. 

Finally, we considered both GUP-induced modifications of the path-fractal in the position space representation and EUP-induced modifications of the path-fractal in the momentum space representation, using the smeared space model \cite{Lake:2018zeg,Lake:2019oaz,Lake:2019nmn,Lake:2021beh,Lake2020-1,Lake2020-2}. 
In $(3+1)$ spacetime dimensions the minimum length was assumed to be of the order of the Planck length, $l_{\rm Pl} = \sqrt{\hbar G/c^3}$, and the minimum momentum was assumed to be of the order of the de Sitter momentum, $m_{\rm dS}c = \hbar\sqrt{\Lambda/3}$, where $\Lambda$ is the cosmological constant, in order to ensure consistency with the gedanken experiment derivations of the GUP and EUP. 

Drawing on our analyses of previous models, we determined the exact form of $D_{\rm H} = D_{\rm H}(\xi,\epsilon)$ and $\tilde{D}_{\rm H} = \tilde{D}_{\rm H}(\xi,\tilde{\epsilon})$, where $\tilde{\epsilon}$ is the ratio of the minimum momentum to the momentum-measurement resolution scale set by the experimental apparatus that is used to resolve the quantum path, $\Delta p$. 
Our main results can be summarised as follows: 

\begin{enumerate} [label=(\roman*)]
	\item The path-fractals in the position and momentum space representations are no longer isomorphic, so that their Hausdorff dimensions are no longer equal, $D_{\rm H} \neq \tilde{D}_{\rm H}$.  
	\item GUP-induced modifications to the position space path-fractal increase $D_{\rm H}$ whereas EUP-induced modifications to the momentum space path-fractal decrease $\tilde{D}_{\rm H}$, relative to their canonical values.
	\item In position space, the extremal case corresponds to setting $\Delta t \rightarrow \infty$ and $\Delta x \simeq l_{\rm Pl}$, giving $D_{\rm H} \simeq 3$. 
	         This is in accordance with our intuition that the additional stochastic motion induced by quantum fluctuations of the background should increase the Hausdorff dimension of the particle path in real space. 
	         The maximum effect is observed when the path is probed at the Planck scale, as expected. 
	\item In momentum space, the extremal case corresponds to setting $\Delta t \rightarrow \infty$ and $\Delta p \simeq m_{\rm dS}c$, giving $\tilde{D}_{\rm H} \simeq 1$. 
	         This appears counter-intuitive, at first, but is the logical corollary of the increase in $D_{\rm H}$. 
	         In this model, the momentum space picture is again dual to the position space representation. 
	         Therefore, an increase in stochastic motion in the latter, relative to canonical QM, results in a decrease in stochastic motion in the former. 
	         We note, however, that  $\tilde{D}_{\rm H} \simeq 1$ does not correspond to the classical regime. The particle path remains embedded as a one-dimensional fractal within the three-dimensional bulk space of the smeared 
	         background. 
	\item The HUP forbids us to realise the extremal cases in the position and momentum space representations, simultaneously, since $\Delta x \gtrsim \hbar/(2\Delta p)$. 
	         Roughly speaking, fractal formation in position space is `bottom up', with self-similarity forming on small scales over short time periods and building up on large scales over longer time-periods.      
	         By contrast, the fractal structure of the particle path in momentum space is formed from the `top down', beginning at large scales over short time-periods and extending to smaller scales at late times. 
	         In canonical QM, the self-similarity of the particle path in position space on scales $(0,\Delta x)$ mirrors that on scales $(\Delta p,\infty)$ but the implementation of different mass/length scales in  
	         the GUP and the EUP, i.e., the Planck scales versus the de Sitter scales, breaks this isomorphism. 
	\item There exists a critical case in which $D_{\rm H}$ and $\tilde{D}_{\rm H}$ are modified symmetrically, such that $D_{\rm H} \simeq 2 + \delta$, $\tilde{D}_{\rm H} \simeq 2 - \delta$, where $\delta^2 \simeq \rho_{\Lambda}/
	         \rho_{\rm Pl}\simeq 10^{-122}$ is the ratio of the dark energy density to the Planck density. 
	         In this case, the energy density of the particle wave function is comparable to $\rho_{\Lambda} \propto m_{\Lambda}^4$, where $m_{\Lambda} \simeq \sqrt{m_{\rm Pl}m_{\rm dS}} \simeq 10^{-3} \, {\rm eV/c^2}$, and the sampling resolutions are chosen such  
	         that $\Delta p \simeq \hbar/\Delta x \simeq \hbar /(c\Delta t) \simeq m_{\Lambda}c$. 
	         Though practically indistinguishable from the canonical quantum regime, the existence of this critical case hints at a deeper connection between the microscopic fractal properties of the dark energy field and its 
	         macroscopic influence on the present-day Universe.  
\end{enumerate}

Finally, we note that the present analysis contains several limitations and could be extended in various ways to provide a more complete picture of minimum-length and minimum-momentum induced modifications to the fractal properties of particle paths. 
In particular, it would be useful to study the Hausdorff and spectral dimensions of paths when the particles are confined within potentials. 
To the best of the authors' knowledge, such as study has not been attempted, even in canonical QM. 
It would be interesting to consider a simple particle-in-a-box, in all three models investigated in this work, and to investigate the limiting scenarios in which the box width tends to $l_{\rm Pl}$ or $l_{\rm dS}$. 

The latter may be seen as a naive model of a quantum particle confined within the de Sitter horizon of the present day Universe, whereas the former represents confinement on the smallest possible scales, corresponding to the horizon of the Universe immediately after the big bang. 
It would be interesting to see how both $D_{\rm H}$ and $\tilde{D}_{\rm H}$ change over cosmic time-scales and how the they are related to the probe time $\Delta t$ at different epochs.   

\acknowledgments

My thanks to the Frankfurt Institute for Advanced Studies, for gracious hospitality during the preparation of this manuscript, and to Piero Nicolini, for helpful discussions and guidance over the course of this project. 
This work was supported by the Natural Science Foundation of Guangdong Province, grant no. 008120251030.


\end{document}